\documentclass[twocolumn,times,tighten]{aastex61}

\newcommand\nH{n_{\rm H}}

\newcommand\NH{N_{\rm H}}
\newcommand\cm{{\rm\ cm}}
\newcommand\pc{{\rm pc}}
\newcommand\msun{{\rm M_\odot}}
\newcommand\XCO{X_{\rm CO}}
\newcommand\XCI{X_{\rm CI}}
\newcommand\XCII{X_{\rm CII}}
\newcommand\WCO{W_{\rm CO}}

\newcommand\Hm{{{\rm H}_2}}
\received{}
\revised{}
\accepted{}
\submitjournal{ApJ}

\usepackage{color,comment}

\usepackage[normalem]{ulem}

\shorttitle{Dark Molecular Gas in Galaxies}
\shortauthors{Qi Li, Narayanan, Dav\'e \& Krumholz.}

\begin{document}

\title{Dark Molecular Gas in Simulations of $z \sim 0$ Disc Galaxies}

\correspondingauthor{Qi Li}
\email{pg3552@ufl.edu}
\author[0000-0001-8015-2298]{Qi Li}
\affil{Department of Astronomy, University of Florida, 211 Bryant Space Sciences Center, Gainesville, FL 32611 USA}
\author[0000-0002-7064-4309]{Desika Narayanan}
\affil{Department of Astronomy, University of Florida, 211 Bryant Space Sciences Center, Gainesville, FL 32611 USA}
\affil{University of Florida Informatics Institute, 432 Newell Drive, Gainesville, FL 32611 USA}
\affil{Cosmic Dawn Center (DAWN), Niels Bohr Institute, University of Copenhagen, Juliane Maries vej 30, DK-2100 Copenhagen, Denmark}
\author[0000-0003-2842-9434]{Romeel Dav\`e}
\affil{Institute for Astronomy, Royal Observatory, Edinburgh, EH9 3HJ, UK}
\affiliation{University of the Western Cape, Bellville, Cape Town 7535, South Africa}
\affiliation{South African Astronomical Observatory, Cape Town 7925, South Africa}
\author[0000-0003-3893-854X]{Mark R. Krumholz}
\affil{Research School of Astronomy and Astrophysics, Australian National University, Canberra 2611, A.C.T., Australia}
\affil{Centre of Excellence for Astronomy in Three Dimensions (ASTRO-3D), Australia}



\begin{abstract}
The $\rm H_2$ mass of molecular clouds has traditionally been traced
by the CO(J=1-0) rotational transition line.  This said, CO is
relatively easily photodissociated, and can also be destroyed by
cosmic rays, thus rendering some fraction of molecular gas to be
``CO-dark''. We investigate the amount and physical properties of
CO-dark gas in two $z \sim 0$ disc galaxies, and develop predictions for
the expected intensities of promising alternative tracers ([C~\textsc{i}] 609
$\mu$m and [C~\textsc{ii}] 158 $\mu$m emission).  We do this by combining
cosmological zoom simulations of disc galaxies with
thermal-radiative-chemical equilibrium interstellar medium (ISM)
calculations to model the predicted H~\textsc{i} and $\rm H_2$ abundances and CO
(J=1-0), [C~\textsc{i}] 609 $\mu$m and [C~\textsc{ii}] 158 $\mu$m emission properties. Our
model treats the ISM as a collection of radially stratified clouds
whose properties are dictated by their volume and column densities, the
gas-phase metallicity, and the interstellar radiation
field and cosmic ray ionization rates.  Our main results follow.
Adopting an observationally motivated definition of CO-dark 
gas, i.e. $\Hm$ gas with $W_{\rm CO} < 0.1 $ K-km/s, we find that a
significant amount ($\ga 50\%$) of the total $\rm H_2$ mass
lies in CO-dark gas, most of which is diffuse gas, poorly
shielded due to low dust column density.  
The CO-dark molecular gas tends to be dominated by [C~\textsc{ii}], though [C~\textsc{i}] also serves as a bright
tracer of the dark gas in many instances.  At the same time, [C~\textsc{ii}]
also tends to trace neutral atomic gas.  As a result, when we quantify the conversion 
factors for the three carbon-based tracers of molecular gas, we find that [C~\textsc{i}] suffers the least contamination from
diffuse atomic gas, and is relatively insensitive to secondary parameters.
\end{abstract}




\section{Introduction}
Star formation as observed in the local Universe occurs exclusively in
giant clouds of molecular hydrogen ($\Hm$)
\citep[e.g.][]{kennicutt12a,krumholz14a,lada03a}.  While $\Hm$ is the 
most abundant constituent of these giant
molecular clouds (GMCs), its low mass requires temperatures of $\sim
500$ K to excite the first quadrapole line.  As a result, direct $\Hm$
emission from the relatively cold (T$\sim 10-30$ K) interstellar
clouds is relatively faint.

With a typical abundance of $\sim 10^{-4} \ \times$ H$_2$
\citep{lee96a}, $^{12}$CO (hereafter, CO) is the second most abundant
molecule in GMCs, and used as a common tracer of the underlying
molecular hydrogen.  The luminosity from the ground rotational state
of CO (CO J=1-0) is typically used to convert to an $\Hm$ mass via a CO-$\Hm$ conversion factor:
\begin{equation}
X_{\rm CO} = \frac{\WCO}{\NH}
\end{equation}
where $W_{\rm CO}$ is the velocity integrated CO intensity (in
K-km/s), and $\NH$ is the $\Hm$ column density.

The value of this CO-$\Hm$ conversion factor is hotly debated
(see, the reviews by \citealt{bolatto13a} and \citealt{casey14a}).
Fundamentally, there are two issues.  First, even at a fixed CO
abundance, the velocity integrated CO intensity $W_{\rm CO}$ depends
on the physical properties (kinetic temperature and velocity
dispersion) of the gas \citep{narayanan11b,narayanan12a}.  These vary
both within and amongst galaxies, and therefore can drive variations
in $X_{\rm CO}$ accordingly
\citep[e.g.][]{feldmann12a,shetty11a,shetty11b}.  Second, CO can be
both photodissociated and destroyed by cosmic rays (CRs) via collisions with He$^+$
\citep[][]{bisbas15a,narayanan17a,gong17a,gong18a}.  Because of this,
there can be molecular gas that is significantly depleted in CO.
Understanding the origin and physical properties of this so-called
``CO-dark molecular gas'' is the main purpose of this investigation.

Indeed, observational studies have uncovered dark molecular gas in the
Milky Way and nearby galaxies.  The principal methods for
characterizing dark gas thus far include the detection of $\Hm$ gas
via $\gamma$-ray observations \citep[e.g.][]{grenier05a}, dust
continuum emission \citep[e.g.][]{planck11a} and C$^+ 158 \mu$m
emission \citep{pineda13a,pineda14a,pineda14b,langer14a}.  
Defining gas with CO intensity weaker than some threshold as CO-dark, these
methods have found that as much as $\sim 30-70\%$ of the molecular gas
in the Galaxy may be in a CO-dark phase.

Potential tracers of CO dark molecular gas include the [C~\textsc{i}] $^3{\rm
  P}_1 \rightarrow ^3{\rm P}_0$ $609 \mu$m line, and the [C~\textsc{ii}] $^3{\rm
  P}_{3/2} \rightarrow ^3{\rm P}_{1/2}$ $158 \mu$m line.  Both C~\textsc{i} and
C$^+$ are natural byproducts of the ultraviolet-induced
photodissociation of CO, or cosmic-ray driven ion-neutral reaction
\citep[e.g.][]{bisbas15a,narayanan17a}.  C$^+$ can be excited by
collisions with a variety of partners, i.e. electrons, H~\textsc{i}, and $\Hm$
\citep{goldsmith12a,herreracamus15a}. Because it can be excited by collisions 
with all these partners, and because the low ionization potential of C (11.3 eV) 
renders C$^+$ the dominant form of carbon in most of the neutral ISM, 
[C~\textsc{ii}] emission can arise from nearly every phase of the ISM. This said, \citet{pineda14a} used Herschel
observations to constrain the origin of the bulk of Galactic [C~\textsc{ii}] emission
as coming from molecular gas.  Similarly, 
\citet{olsen15a} suggested (via numerical simulations) that this is
likely true even for $z \sim 2$ main sequence galaxies.  Even so, at
lower metallicities, the fraction of [C~\textsc{ii}] emission that originates
from ionized gas may increase \citep{olsen17a}.

Similarly, the [C~\textsc{i}] $^3{\rm P}_1 \rightarrow ^3{\rm P}_0$ transition
line has an excitation potential $\sim 23.6$ K , and can therefore be
excited in typical molecular clouds. Observational studies have shown
spatial correlation with both low-J CO emission and $\Hm$ abundances
\citep{papadopoulos04b,bell07,walter11,israel15}. \citet{bothwell17}
show similar [C~\textsc{i}] and CO linewidths of a range of high-$z$ galaxies.
Those observations span a wide range of environments, from local
molecular clouds to high redshift (U)LIRGS, implying that [C~\textsc{i}] may be
able to trace the gas component traditionally traced by CO emission.

The results from theoretical models are mixed.  With respect to [C~\textsc{ii}],
theoretical studies have typically focused on disentangling different
components of [C~\textsc{ii}] emission.  Utilizing hydrodynamics simulations of
star formation regions combined with photodissociation region (PDR) and
radiative transfer modeling, \citet{accurso17} find that
$\sim60$--$80\%$ of [C~\textsc{ii}] comes from molecular regions, depending
mainly on the specific star formation rate. Similarly, \citet{olsen17a}
found $\sim 66\%$ of [C~\textsc{ii}] comes from molecular gas in star forming
galaxies at $z\sim6$. Using a kpc-scale patch extracted from 
isolated galaxies, \citet{glover16a} found a weak correlation between 
the [C~\textsc{ii}] intensity and $\NH$.

On the other hand, [C~\textsc{i}] has been historically interpreted as only a
tracer of surface layer of a thin PDR layer on the surface of clouds
\citep[e.g.][]{tielens85}.  This interpretation was questioned in
subsequent studies.  For example, \citet[][]{papadopoulos04a}
suggested that [C~\textsc{i}] can be widespread in molecular regions
traditionally traced by CO emission, due to increased exposure to
interstellar radiation fields from turbulence
\citep{xie95,spaans97,cubick08}.  

A number of groups have investigated the behavior of [C~\textsc{i}] in smaller scale molecular cloud simulations 
\citep{offner14a,glover15a}. They point out
several advantages of C~\textsc{i} over CO: the column density regime of [C~\textsc{i}]
where the corresponding conversion factor $X_{\rm C~\textsc{i}}$ remains approximately
constant is larger, and $X_{\rm C~\textsc{i}}$ is less sensitive to secondary parameters
such as interstellar radiation field strength. 

Thus far, what has been missing is an investigation into the expected
fraction of CO-dark molecular gas on galaxy-wide scales.
Understanding this is the focus of the present paper.  To do this, we
take advantage of a combination of cosmological zoom-in simulations
with thermal-radiative-chemical equilibrium interstellar medium (ISM)
calculations to model the predicted $\Hm$, H~\textsc{i}, CO(1-0), [C~\textsc{i}] and [C~\textsc{ii}]
emission\footnote{Because we refer to both abundances and intensities
  in this paper, we must be a bit careful with our terminology.  We
  hereafter refer to CO (J=1-0) emission as CO (1-0), while we refer
  to the molecule itself as ``CO''.  We refer to [C~\textsc{i}] $^3$P$_1
  \rightarrow ^3$P$_0$ emission as [C~\textsc{i}], while we refer to the neutral
  atom itself as ``C~\textsc{i}''.  Finally, we refer to [C~\textsc{ii}] $158 \mu$m
  emission as [C~\textsc{ii}], while we refer to singly ionized carbon as
  C$^+$.} from giant clouds in galaxy simulations.  Here, we focus on
$z \sim 0$ disc galaxies, though we plan to extend our models to galaxy
populations at high-redshift in subsequent work.  In this paper, we
ask three fundamental questions:
\begin{enumerate}
\item How much CO-dark molecular gas is there in $z \sim 0$ disc Galaxies?
\item What are the physical properties of this dark molecular gas?
  \item What are the best alternative tracers of CO-dark gas?
\end{enumerate}

In \S \ref{sec:m}, we summarize our simulation methods. 
In \S \ref{sec:i}, we show sample results from simplified cloud based on our modeling to develop our physical intuition.
In \S \ref{section:results}, we investigate the amount and properties of CO-dark molecular gas, and examine
the utility of different carbon-based tracers. In \S \ref{sec:d}, we then discuss
our results in the context of other relevant theoretical studies and the
sensitivity of uncertain parameters in our model. We summarize
in \S \ref{sec:c}.
 
\section{Methods}
\label{sec:m}

\subsection{Cosmological Zoom Galaxy Formation Simulations}

We examine two Milky Way-like galaxies formed in cosmological zoom-in
simulations using the {\sc mufasa} physics suite
\citep{dave16a,dave17b,dave17a}.  The basic physics in these
simulations is described in \citet{narayanan18a,narayanan18b} and
\citet{privon18a}, and we refer the reader to these papers for
details, though summarize the salient points here.

We first simulate a $50$ Mpc$^3$ dark matter only box at relatively
low mass resolution ($M_{\rm DM} = 7.8 \times 10^8 h^{-1}$ M$_\odot$)
down to redshift $z=0$. We conduct this (and our main zoom in
simulation) with the hydrodynamic code {\sc gizmo} in Meshless Finite
Mass (MFM) mode \citep{hopkins15a,hopkins17a}.  We employ the cubic
spline kernel with $64$ neighbors, which leads to a minimum 
smoothing length $\epsilon \sim 30$~pc. We evaluate volume and column 
densities of gas on the scale of the smoothing length. Our initial conditions are
generated with {\sc music} \citep{hahn11a}, and are exactly the same a
the {\sc mufasa} cosmological hydrodynamic simulation series
\citep{dave16a,dave17b,dave17a}.

From the $z=0$ snapshot, we identify dark matter halos with {\sc
  caesar} \citep{thompson14a}, and select $2$ halos with dark matter
halo masses $M_{\rm halo}$ = $1.9 \times 10^{12} M_\odot$ and $2.1
\times 10^{12} M_\odot$. These are named ``Halo 352'' and ``Halo 401''
respectively, hereafter.  We identify all particles within $2.5 \times
r_{\rm max}$, where $r_{\rm max}$ is the distance of the particle with
maximum radius from the halo center of mass at $z=0$.  We track these
particles back to redshift $z=249$, and split them in order to achieve
higher mass resolution.  At this point, we also add baryons
according to the cosmic baryon fraction.  Our final particle masses
are $M_{\rm DM} = 1 \times 10^{6} h^{-1}$ M$_\odot$, and baryon mass
$M_{\rm b} = 1.9 \times 10^5 h^{-1} $M$_\odot$.

We re-run these higher-resolution simulations from an initial redshift
$z=249$ down to $z=0$. Our galaxy formation physics follows that
employed in the {\sc mufasa} cosmological hydrodynamic simulation.  We
refer the reader to \citet{dave16a} for a detailed description of
these models.  Briefly, we use the {\sc GRACKLE}-3.1 chemistry and cooling library \citep{smith17a}, which includes primordial and metal-line cooling. Gas is allowed to cool down to $10^4$~K, below which it is pressurized in order to resolve Jeans mass. Stars form in molecular gas, where the H$_2$
fraction is calculated following the \citet{krumholz09a} methodology.
We impose a minimum value of metallicity $Z = 10^{-3}
  Z_\odot$ entering the calculation. 
  We take this floor value from \citet{kuhlen12a}, who show that a minimum metallicity of this order should be 
  produced by metal enrichment from population III stars, which form in early halos with masses well below 
  the resolution limit of our simulations. The rate of star formation proceeds following a
  volumetric \citet{schmidt59a} relation, with an enforced efficiency
  per free fall time of $\epsilon_{\rm ff} = 0.02$ as motivated by
  observations
  \citep[e.g.][]{kennicutt98a,kennicutt12a,krumholz12a,narayanan12a,vutisalchavakul16a,heyer16a,leroy17a}.

  These stars drive winds in the interstellar medium.  This form of
  feedback is modeled as a two-phase decoupled wind.  The modeled
  winds have an ejection probability that scales with both the SFR and
  the galaxy circular velocity (a quantity calculated on the fly via
  fast friends-of-friends galaxy identification).  The nature of these
  scaling relations follow the results from higher-resolution studies
  in the Feedback In Realistic Environments zoom simulation campaign
  \citep[e.g.][]{muratov15a,hopkins14b,hopkins17b}.  We additionally
  include feedback from longer-lived stars (e.g. asymptotic giant
  branch stars and Type 1a supernovae) following \citet{bruzual03a}
  stellar evolution tracks with a \citet{chabrier03a} initial mass
  function. MUFASA also includes an additional, artificial "quenching 
  feedback" intended to suppress star formation in massive halos, but we 
  do not use it in the simulations presented in this paper because our halos 
  are below mass threshold where it applies.
  
  Metal enrichment yields for type Ia and type II supernovae are drawn
  from \citet{iwamoto99a} and \citet{nomoto06a}, respectively
  \citep[though note that the latter yields are reduced by a factor
    $50\%$ to match the mass-metallicity relation statistically, following the discussion in][]{dave16a}.  Asymptotic Giant
  Branch (AGB) star yields are drawn from \citet{oppenheimer06a}.

\begin{figure*}[]
\centering
\includegraphics[scale=0.7]{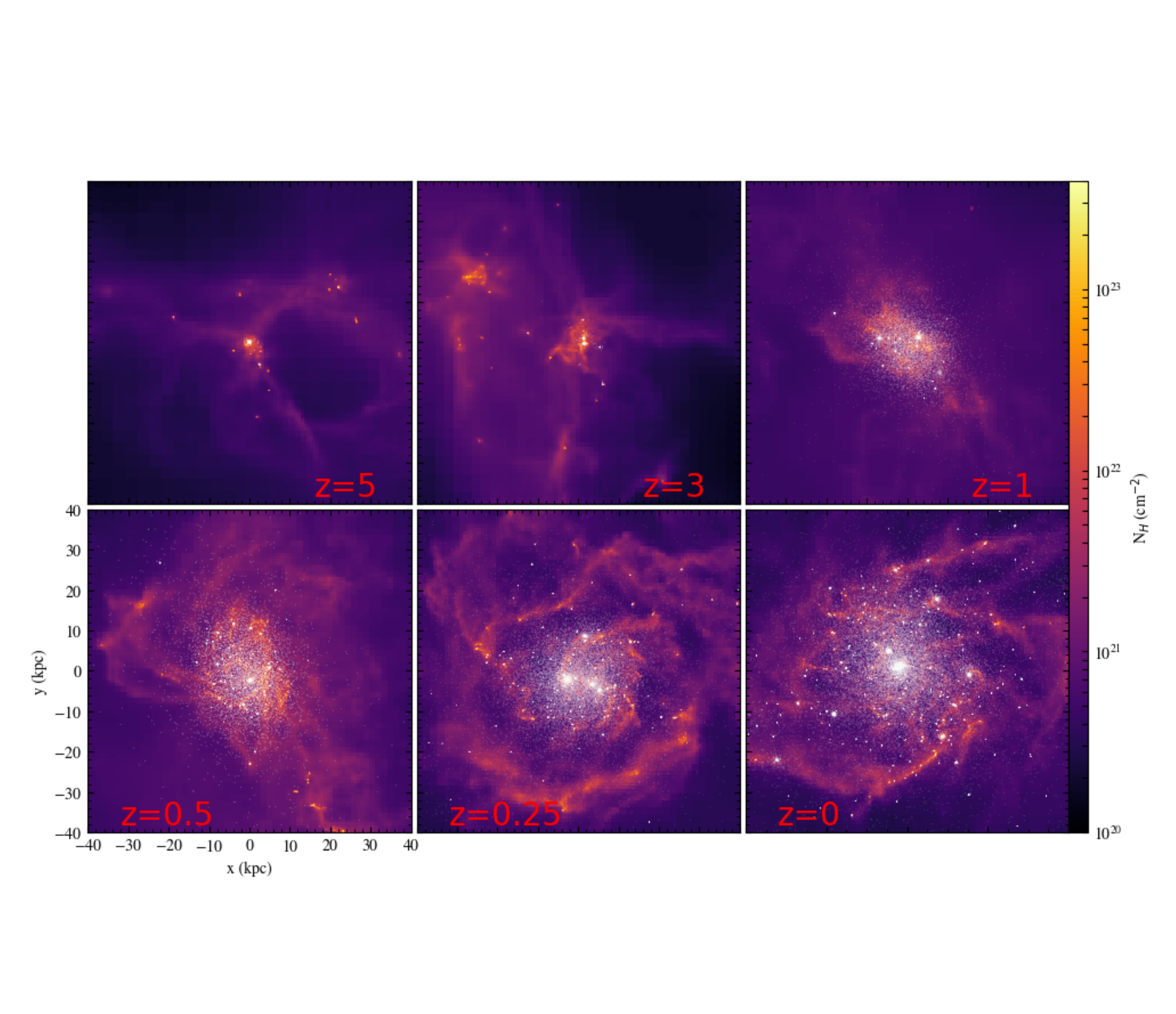}
\vspace{-0.5cm}
\caption{Gas surface density projection maps of in formation from redshifts $z=6$ to 0.  
The panels are $80$ kpc (comoving) on a side, and the
  color scale is common for all images with the color scale on the
  right. The star particles are overplotted to show the stellar disks.
\label{figure:morphology}}
\end{figure*}

\begin{figure}[]
\centering
\includegraphics[width = 0.5 \textwidth]{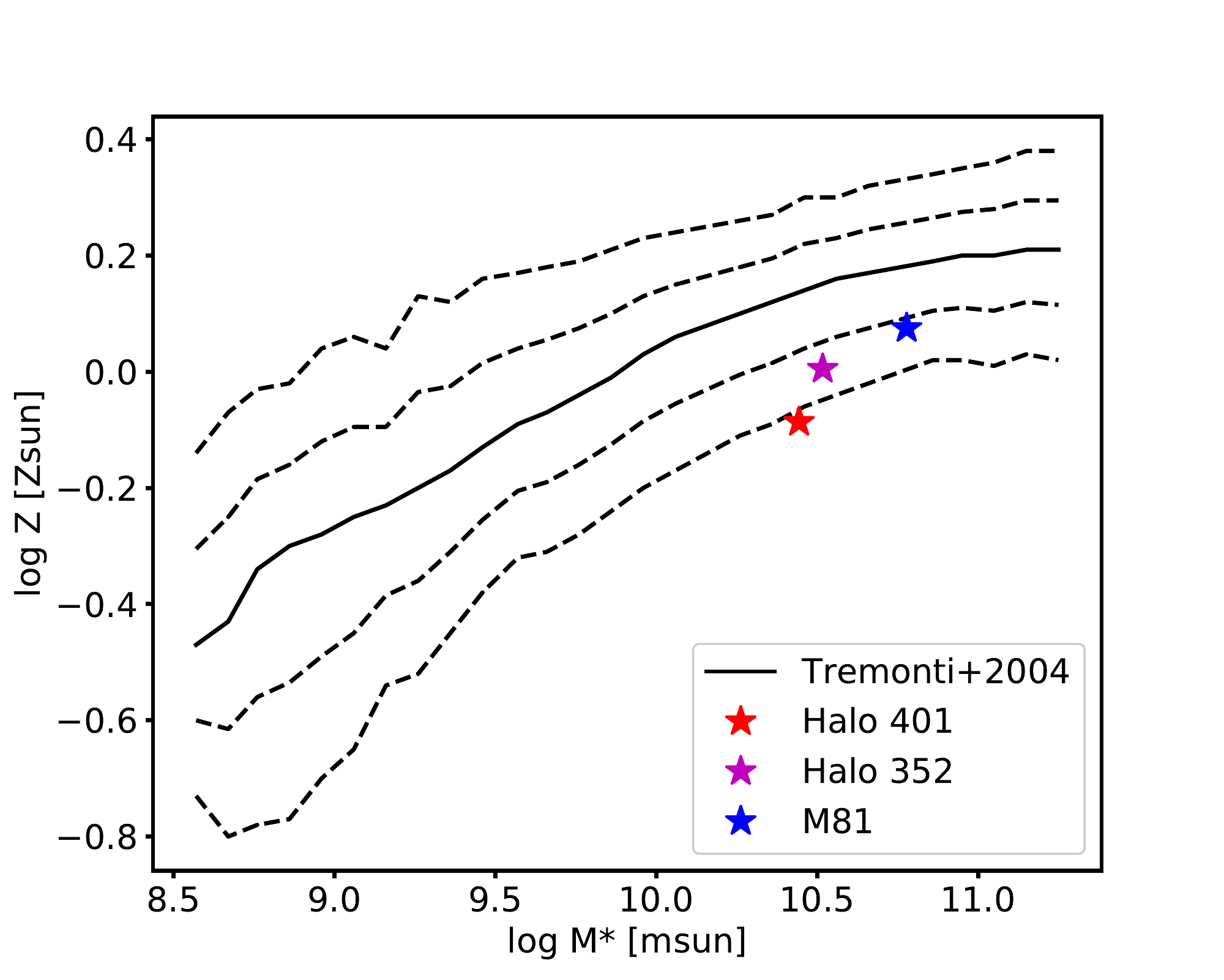}
\caption{The mass-metallicity relation (solid line, with dashed lines
  representing $1\sigma$ and $2\sigma$ uncertainties;
  \citealt{tremonti04a}). The red stars denote the mass-weighted
  metallicity of molecular gas versus stellar mass in our simulated
  galaxies (Halo 401 and Halo 352). The blue star denotes M81 \citep{kudritzki12a}}.
\label{fig:rg1}
\end{figure}

 The results of these simulations are $2$ disc-like galaxies at
 redshift $z=0$.  In Figure~\ref{figure:morphology}, we show the gas
 surface density images of one of our model galaxies from $z=5$ to 0, and list the final physical properties of these galaxies
 in Table~\ref{table:simsum}. Figure~\ref{fig:rg1} shows where the simulated galaxies 
 are located in the mass-metallicity plane. Compared against the mass-metallicity relation 
 (\citealt{tremonti04a}), our galaxies have only slightly lower gas metallicity, which could 
 lead to a smaller amount of CO-bright gas.
  \citet{abruzzo18a} show the location
 of these galaxies on both the SFR-$M_*$ relation, as well as the
 $M_*$-$M_{\rm halo}$ relation.
 
  \begin{deluxetable}{lcccc}
\tablecaption{Zoom Simulation Summary}
\tablehead{\colhead{Halo ID} & \colhead{M$_{\rm DM}$($z=0$)} & \colhead{M$_{*}$($z=0$)} & \colhead{M$_{\rm g,HI+H_2}$($z=0$)} & \colhead{SFR}\\
\colhead{} & \colhead{$M_\odot$} & \colhead{$M_\odot$} & \colhead{$M_\odot$} &\colhead{$M_\odot$ yr$^{-1}$}}  
\startdata
mz352 & $1.9\times 10^{12}$ & $3.3 \times 10^{10}$ & $1.3 \times 10^{10}$ & $3.8$\\
mz401 & $2.1\times 10^{12}$ & $2.6 \times 10^{10}$ & $4.2 \times 10^{10}$ & $4.3$\\
\enddata
\tablecomments{Halo ID number, halo mass, stellar mass, neutral gas (H~\textsc{i} + H$_2$) mass and SFR.  All physical quantities are calculated at $z=0$}
\label{table:simsum}
\end{deluxetable}

\subsection{Computation of Thermal States, Chemical States and Line Emission}
We calculate the chemical abundances and line emission of the gas
through post-processing the gas particles from the simulated
galaxy. We utilize the code Derive the Energetics and Spectra of
Optically Thick Interstellar Clouds ({\sc Despotic}; \citealt{krumholz14a})
to compute the chemical state, temperature and level populations in
each layer. We briefly summarize the computation here, and refer
readers interested in further details to \citet{krumholz14a} and
\citet{narayanan17a}.

The computation of emergent luminosity requires three steps:
calculating thermal equilibrium, chemical equilibrium, and statistical
equilibrium.  In practice, these three steps are done
iteratively.  To perform these, we first assume that every particle is
a sphere, and is radially stratified into $8$ concentric zones. 
Note, we test the convergence by varying the number of zones from 8 to 64, 
and show in Appendix that the $8$-zone model is sufficient to produce a converged result.
The equilibrium calculations for each zone is done independently from
one another.

Each particle has a density $\rho$, and particle mass $M$ known from
the hydrodynamic simulations.  We assign the column density
of H nuclei $N_H = \left(3/4\right) \Sigma_g/\mu_H$ where $\mu_H$ is
the mean molecular weight.  The factor $3/4$ is the difference
between mean column density and center-to-edge column density.  Within
each zone, we adopt total abundances of [C/H] = $2 \times 10^{-4}
Z'$ consistent with their Milky Way abundances
\citep{lee96a,draine11a}.  Here, $Z'$ is the metallicity scaled to a
solar value of 0.0134 \citep{asplund2009}.

We model the chemical state of each zone using the reduced
carbon-oxygen chemical network as developed by \citet{nelson99a},
combined with the \citet{glover07a} non-equilibrium hydrogen chemical
network (hereafter NL99+GC).  The reaction rates and full network are described in detail
in Table 2 of \citet{narayanan17a}.  This network specifies the
fraction of carbon in CO, C~\textsc{i} and C$^+$ in each zone, as well as H~\textsc{i} and
H$_2$.  

This network requires that we specify the strength of the unshielded
interstellar radiation field (ISRF), which we characterize in terms of the
far-ultraviolet radiation field.  We normalize this field to the solar
neighborhood value $\chi_{\rm FUV}$, which we assume is linearly
proportional to the total galaxy SFR.  The normalization of this scale
is set to $1 M_\odot$ yr$^{-1}$, based on the approximate observed Milky Way value
\citep{robitaille10a,chomiuk11a}.

We additionally must specify the amount by which we reduce the rates
of all photochemical reactions by dust shielding in the interiors of
clouds.  We characterize this dust-shielding in terms of the visual
extinction $A_{\rm v} = \left(1/2\right)\left(A_{\rm v}/N_H\right)N_H$, where the
ratio of $\left(A_{\rm v}/N_H\right)$ is the dust extinction per H nucleus
at $V$ band, and the factor of $1/2$ represents the rough average
column over the entire volume of the cloud.  The reduction in the
H$_2$ dissociation rate is evaluated using the \citet{draine96a}
shielding function, and the reduction in CO photodissociation is
determined using an interpolated version of the shielding function of
\citet{vandishoeck88a}.  

We further specify a cosmic ray (CR) ionization rate.  Recent observations suggest
the CR ionization rate $\xi \sim 10^{-16}$ s$^{-1}$ in the diffuse ISM
\citep{neufeld10a,indriolo15a,neufeld17a}.  However, considering that
a significant amount of the CR flux is at low energies, the
ionization rate in the interiors of molecular clouds will likely be lower due
to shielding.  Similarly, the relatively large values of $\xi \sim
10^{-16}$ s$^{-1}$ are at tension with the low inferred temperatures
of typical molecular clouds \citep[e.g.][]{narayanan11b,narayanan12b}.
Because of this, we adopt a more conservative value of $\xi =
10^{-17}$ s$^{-1}$. We assume that the CR ionization rate
$\xi$ scales linearly with the SFR, i.e. $\xi = 10^{-17}$
SFR/$M_\odot\ \mathrm{yr}^{-1}$ s$^{-1}$.  Note, increasing this value will have the effect of increasing the CO-dark fraction in clouds.

Alongside our chemical equilibrium calculations, we determine the
thermal state of each zone of our model clouds.  To do this, we follow
\citet{goldsmith01a} and \citet{krumholz11a} in balancing the relevant heating and
cooling processes.  The heating processes we consider are the cosmic
microwave background (CMB) (set to $T_{\rm CMB} = 2.73 $ K), heating
of the dust by the ISRF, heating of the dust by a background thermal infrared 
field with an effective temperature of 10K, grain photoelectric heating in the gas, 
and cosmic ray heating of the gas. The cooling processes are dominated by line cooling of
the gas by C$^+$, C~\textsc{i}, O and CO, as well as of atomic hydrogen excited
by electrons via the Ly$\alpha$ and Ly$\beta$ lines (processes that
can be relevant at $T \ga 5000$ K, which is possible in the outskirts
of our clouds).  There is additionally an energy exchange between gas
and dust.

Finally statistical equilibrium is assumed to determine the level
populations of each species.  {\sc Despotic} uses the escape
probability approximation for the radiative transfer problem, where
$f_i$, the fraction of each species in quantum state $i$ is computed
solving the linear system of equations
\begin{eqnarray}
\lefteqn{\sum_j f_j \left[q_{ji} + \beta_{ji} (1 + n_{\gamma,ji}) A_{ji} + 
\beta_{ij} \frac{g_i}{g_j} n_{\gamma,ij} A_{ij} \right]}
\nonumber
\\
& = & f_i \sum_k \left[q_{ik} + \beta_{ik} (1 + n_{\gamma,ik}) A_{ik} + 
\beta_{ki} \frac{g_k}{g_i} n_{\gamma,k i} A_{ki}\right]\qquad
\label{equation:kmt_stateq}
\end{eqnarray}
where $\sum_i f_i = 1$, $A_{ij}$ are the usual Einstein coefficients
for spontaneous emission (from $i\rightarrow j$), $g_i,g_j$ are
the degeneracies of the states, 
\begin{equation}
n_{\gamma,ij} = \frac{1}{\exp(\Delta E_{ij}/k_B T_{\rm CMB})-1}
\end{equation}
is the photon occupation number of the CMB, $E_{ij}$ is the energy difference
between states $i$ and $j$, and $\beta_{ij}$ is the escape probability
for photons of this energy \citep{draine11a}.  $q_{ij}$ are the
collisional transition rates between states, where the calculation of collision 
rates includes both H~\textsc{i} and H$_2$, in proportion to their abundances 
as determined by the chemical network.  The Einstein and collisional rate 
coefficients all come from the Leiden Atomic and Molecular Data Base \citep[LAMADA;][]
{schoier05a,jankowski05a,wernli06a,yang10a,launay77a,johnson87a,roueff90a,
schroder91a,staemmler91a,barinovs05a,wilson02a,flower77a,flower88a,flower01a,lique13a,wiesenfeld14a}.  

To compute the chemical and thermal states, alongside the emergent
line luminosity, we first guess an initial temperature, chemical state
and level populations for each layer in the cloud.  We then iterate
each process in each zone independently.  The outermost loop is the
chemical network, and this is run to convergence while holding the
temperature fixed.  The middle loop computes the temperature, holding
the level populations fixed.  Finally, the innermost loop (the level
populations) are iterated for each species to convergence.  We iterate
in this manner until the chemical abundances, temperatures, and level
populations are converged.  With these in hand, the total line
luminosity per unit mass in each zone is summed.

Finally, we note that this process is computationally expensive.  When
considering the relatively large number of gas particles in a single
snapshot, it is prohibitive to determine the line luminosity in this
manner on a particle-by-particle basis.  Because of this, we have
built a 4-dimensional lookup table where we compute the aforementioned
{\sc Despotic} calculations in a fine grid over a range of $\nH$,
$\NH$, star formation rates and metallicity.  The lookup table spans
the metallicity, column density, volume density, and SFR space of $0.1
\leqslant Z' \leqslant 1.5$, $1 \leqslant\nH\leqslant 10^4$, $1
\leqslant \NH/(\msun/{\rm pc^2}) \leqslant 10^4$, and $1\leqslant$
SFR/($\msun$/yr) $\leqslant 10^3$.  The spacing of the grid is 0.28,
0.2 dex, 0.2 dex, 0.15 dex for $Z'$, $\nH$, $\NH$ and SFR,
respectively.  The results presented here are all converged with
respect to the grid resolution of the lookup tables.

\section{Developing Intuition: Results from Simplified Cloud Models}
\label{sec:i}
Before we explore our main results, it is worth first going through a
pedagogical exploration of the chemical properties of individual
spherical, radially stratified {\sc Despotic} clouds.  These numerical experiments
will allow us to develop a physical intuition that will be useful when
examining the CO, C~\textsc{i} and C$^+$ abundances in bona fide galaxy zoom
simulations.

\begin{figure}[]
\centering
\includegraphics[width = 0.45 \textwidth]{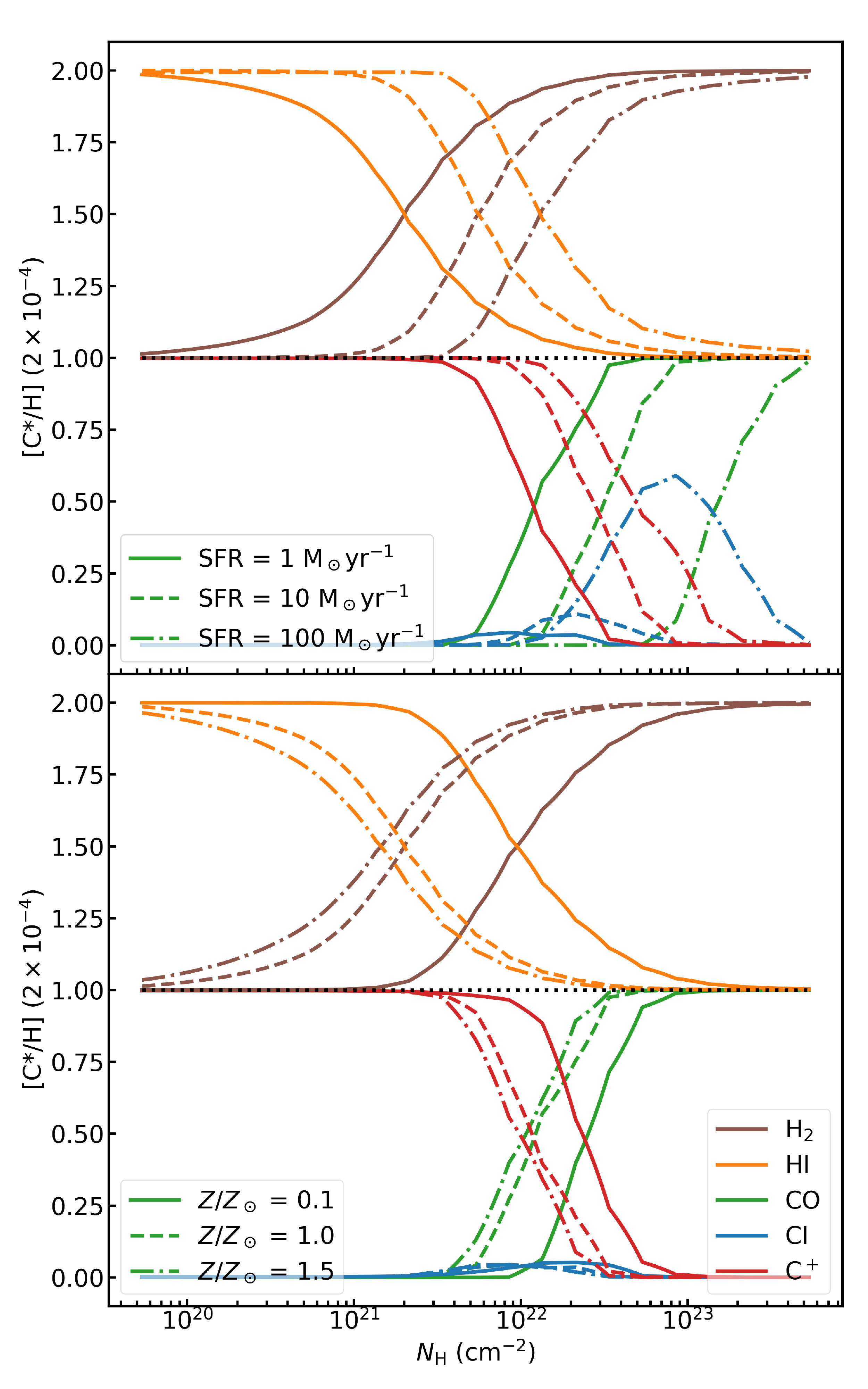}
\caption{Carbon species' abundances from idealized cloud models with a fixed 
density $n_{\rm H} = 100\cm^{-3}$. 
Upper panel: [C*/H] as functions of $\NH$.  We fix the metallicity 
to solar. We vary the star-formation-rate to values of  SFR$=1$ 
(solid line), $10$ (dashed line), $100$ $\msun$/yr (dotted dashed line).
Lower panel: CO (green), 
C~\textsc{i} (blue), C$^{+}$ (red) abundances with regard to H ([C*/H]) 
as functions of $\NH$. We fix the star formation rate to SFR$=1$ $\msun$/yr. 
We vary the metallicity to values of $Z'$= $Z/Z_\odot$ = $0.1$ 
(solid line), $1.0$ (dashed line), $1.5$ (dotted dashed line). 
We overplot the hydrogen species' abundances which are normalized to unity 
instead of $2 \times 10^{-4}$ and a +1 offset.} 
\label{fig:p2}
\end{figure}

For the purpose of the idealized numerical experiments we develop in
this section, we fix the initial density of spherical clouds to $\nH =
100\cm^{-3}$ and run the chemical--thermal--statistical network to
equilibrium.  In Figure~\ref{fig:p2}, we examine the relationship
between the CO, C~\textsc{i} and C$^+$ abundances as a function of column density
when varying the star formation rate (SFR; upper panel) and metallicity (lower panel). 
We overplot the $\Hm$ and H~\textsc{i} abundances which are normalized
to unity instead of $2 \times 10^{-4}$ and a $+1$ offset to show how H$_2$ fractional 
abundances are impacted by SFR and metallicity.

With fixed SFR and metallicity, the H$_2$ abundance increases as the column density
increases, mainly due to the increased ability of hydrogen to
self-shield against the photodissociating radiation. At a fixed
column density, the H$_2$ abundance also increases as the metallicity
increases, owing to the increased efficacy of dust
shielding. Conversely, a higher SFR (which results in a stronger
photo-dissociating FUV field and cosmic-ray ionization rate) tends to
reduce the $\Hm$ abundance.

For a fixed set of physical conditions
(i.e. SFR and $Z'$), C$^+$ tends to dominate the low column density
regime (here, the outskirts of our spherical clouds) while CO
dominates the high column density regime.  As we show in
the upper panel of Figure~\ref{fig:p2}, at a fixed SFR, carbon tends to transition from
ionized form to molecular with increasing metallicity due to
enhanced dust shielding.

Compared to the hydrogen abundances, the transition of carbon species' 
composition occurs at higher column densities due to the lack of self-shielding.
The SFR has a much more significant effect on the abundances of carbon 
in different phases. Enhanced SFRs slightly expand the 
C$^{+}$-dominated region while significantly enlarging the C~\textsc{i}-dominated region. 
This is mainly due to the fact that CO, unlike $\Hm$, is
unable to self-shield.  Its survival is therefore more sensitive to
FUV photo-dissociation, as well as destruction via cosmic rays
\citep[via a two-body reaction from ionized
  He$^+$,][]{bisbas15a,narayanan17a}.  

\section{CO-dark Gas in Cosmological Galaxy Formation Simulations}
\label{section:results}
We now turn our attention to understanding the utility of different
carbon-based tracers of molecular gas in the cosmological zoom
simulations of Milky Way analogs. We frame these results in terms of
answering a number of specific questions of interest.  We assume for
both models an SFR of $1 M_\odot$/yr for snapshots analyzed in this
paper.  This is comparable to the actual SFRs, though it is cleaner when comparing results across multiple models to impose a constant SFR. 

\section{Definitions}
\label{sec:def}
We begin the analysis of our results with some definitions.
Motivated by observations, we define CO-dark gas as 
$\Hm$ gas \textit{layer} with $W_{\rm CO} < 0.1 $ K-km/s. This value
matches the typical sensitivity in observations of nearby molecular
clouds \citep[e.g.][]{pineda08a,smith12a,ripple13a,leroy16a}.

When examining properties of CO-dark gas and the utility of different tracers, 
we only consider gas with $W_{\rm CO} > 10^{-5} $ K-km/s.
As will be shown later in Figure~\ref{fig:t1}, only $\sim 10\%$ of $\Hm$ mass associated 
with gas fainter than this limit, which is extremely diffuse, atomic or ionized. 
In this type of gas, $\Hm$ can only exist sparsely in dense substructures well below the model resolution 
(even the finest resolution of observations). As a result, the emission calculated from our model, 
which is an average of a whole layer, cannot represent the emission from real $\Hm$ regions, 
and is therefore not interesting to discuss.

Because our radially stratified clouds are comprised of a mixture of
H~\textsc{i} and $\Hm$, to simplify our presentation of results, we refer to layers of our clouds as
``molecular'' if $M_{\rm H_2}/M_{\rm HI+H_2} \geq 0.5$ and ``atomic'' if $M_{\rm H_2}/M_{\rm HI+H_2} < 0.5$.

\subsection{Where Does CO(1-0), [C~\textsc{i}] and [C~\textsc{ii}] Emission Come from?}

We first examine what fractions of emission of CO(1-0), [C~\textsc{i}] and [C~\textsc{ii}]
come from molecular versus atomic gas.

\begin{figure*}[]
\centering
\includegraphics[width = 1.0 \textwidth]{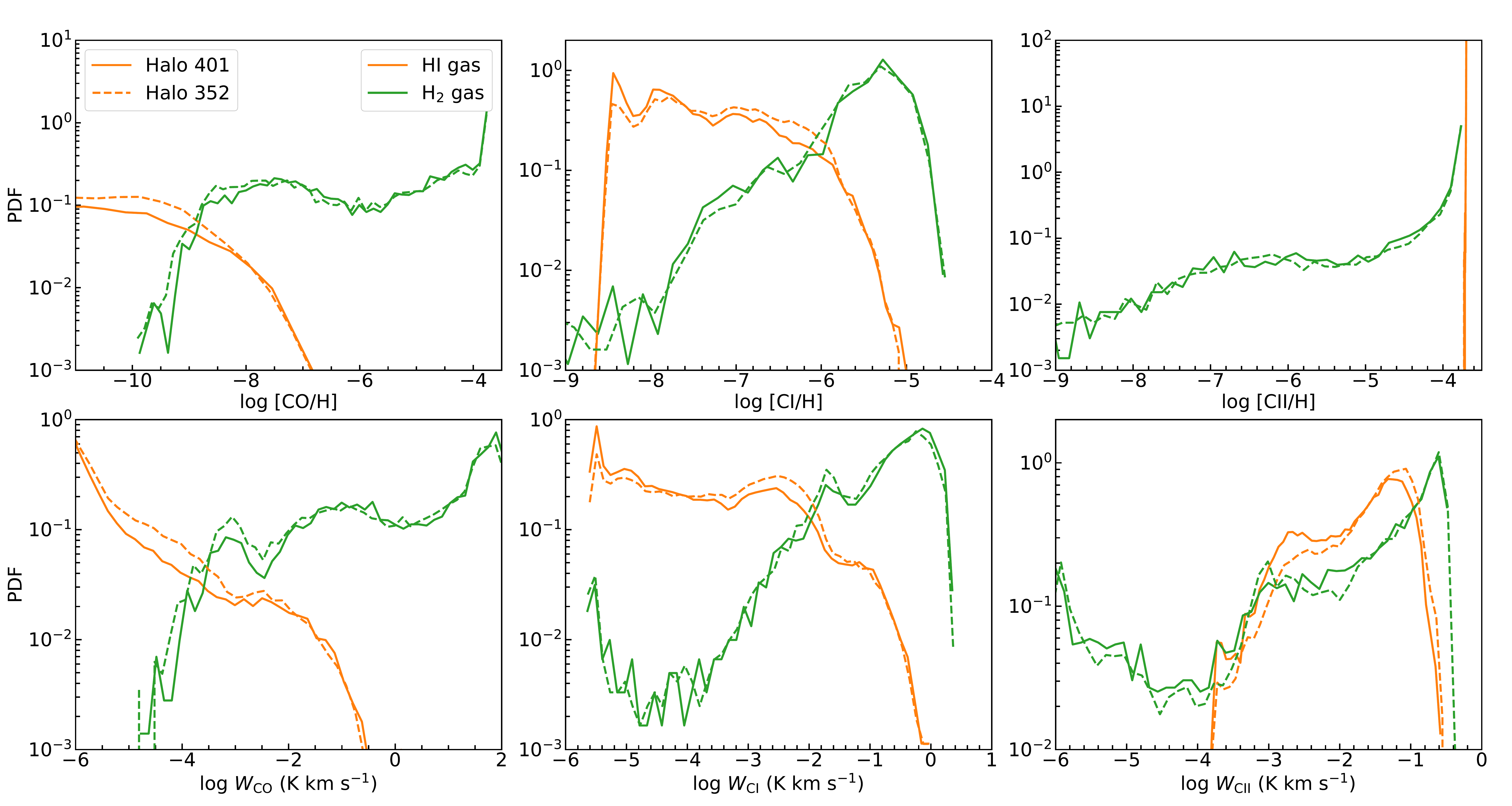}
\caption{Upper row: PDFs of CO, C~\textsc{i} and C$^+$ abundances (from left to
  right) in molecular gas ($\Hm$ abundance $>$ 0.5; green) and atomic
  gas (HI abundance $>$ 0.5; orange) for Halo 401 and Halo 352 at
  redshift $z=0$.  CO does not exist in any appreciable abundance in
  atomic gas, nor does C~\textsc{i}.  Rather, both principally reside in
  molecular $\Hm$ gas.  On the other hand, C$^+$ can reside in both
  atomic and molecular gas, and carbon in atomic gas is almost
  exclusively in the ionized C$^+$ phase, as is shown by the extremely sharp orange PDF in the rightmost panel. Lower row: PDFs of
  velocity-integrated CO(1-0), [C~\textsc{i}], [C~\textsc{ii}] (from left to right)
  intensities (K-km/s) in molecular and atomic gas.  As we transition from molecular (CO 1-0) to atomic ([C~\textsc{i}]) to
ionized ([C~\textsc{ii}]) carbon emission, the fraction of emission that
originates in molecular gas decreases, and the fraction that
originates in atomic gas increases.}
\label{fig:t3-1}
\end{figure*}

In the upper row of Figure~\ref{fig:t3-1}, we show three panels, each
showing the probability distribution functions (PDFs) of CO, C~\textsc{i} and
C$^{+}$ (left to right) abundances for our two model
galaxies at redshift $z=0$. For both galaxies, CO does not
exist in any appreciable abundance in atomic gas, nor does C~\textsc{i}.
Rather, both principally reside in molecular $\Hm$ gas. On the other
hand, C$^+$ can reside in both atomic and molecular gas, and the
carbon in atomic gas is almost exclusively in the ionized C$^+$ phase.

Correspondingly, in the lower row of Figure~\ref{fig:t3-1}, we show
three panels, each showing the PDFs of the intensity from CO(1-0), [C~\textsc{i}]
and [C~\textsc{ii}] (left to right) coming from both atomic and molecular
gas. The orange line in each panel shows the distribution of line
intensities (K-km/s) that come from atomic gas, while the green line
shows the distribution of line intensities that come from molecular
gas.  As we transition from molecular (CO 1-0) to atomic ([C~\textsc{i}]) to
ionized ([C~\textsc{ii}]) carbon emission, the fraction of emission that
originates in molecular gas decreases, and the fraction that
originates in atomic gas increases.  

We quantify the mass
fraction of molecular gas dominated by CO (1-0), [C~\textsc{i}] and [C~\textsc{ii}].  For
example, CO (1-0) is classified as dominating when the CO (1-0) intensity 
(in erg~s$^{-1}$~cm$^{-2}$~sr$^{-1}$ is
greater than either [C~\textsc{i}] or [C~\textsc{ii}]. Nearly $80\%$ of the
molecular gas is dominated by [C~\textsc{ii}], while the bulk of the remainder
is dominated by CO (1-0).  The fraction of molecular gas dominated by
[C~\textsc{i}] emission is negligible; this owes to the relatively narrow set of
physical conditions in which [C~\textsc{i}] emission peaks (c.f. Figure~\ref{fig:p2}).
This said, this should not be interpreted as [C~\textsc{i}] being an ineffective
tracer of molecular gas.  Rather, 
even if [C~\textsc{i}] is mostly fainter than [C~\textsc{ii}], it is still brighter 
than CO(1-0) in the CO-dark gas. In fact, as we will show in \S~\ref{sec:t}, [C~\textsc{i}] 
emission is generally strong enough and can serve as a reasonably reliable tracer of $\Hm$ gas.

\subsection{What Are the Physical Properties of CO-dark Gas?}
\label{sec:p}

Having established that a significant fraction of the molecular gas 
in $z \sim 0$ disc galaxies is CO-dark, we next investigate the physical 
properties of this gas.  

\begin{figure}[]
\includegraphics[width = 0.5 \textwidth]{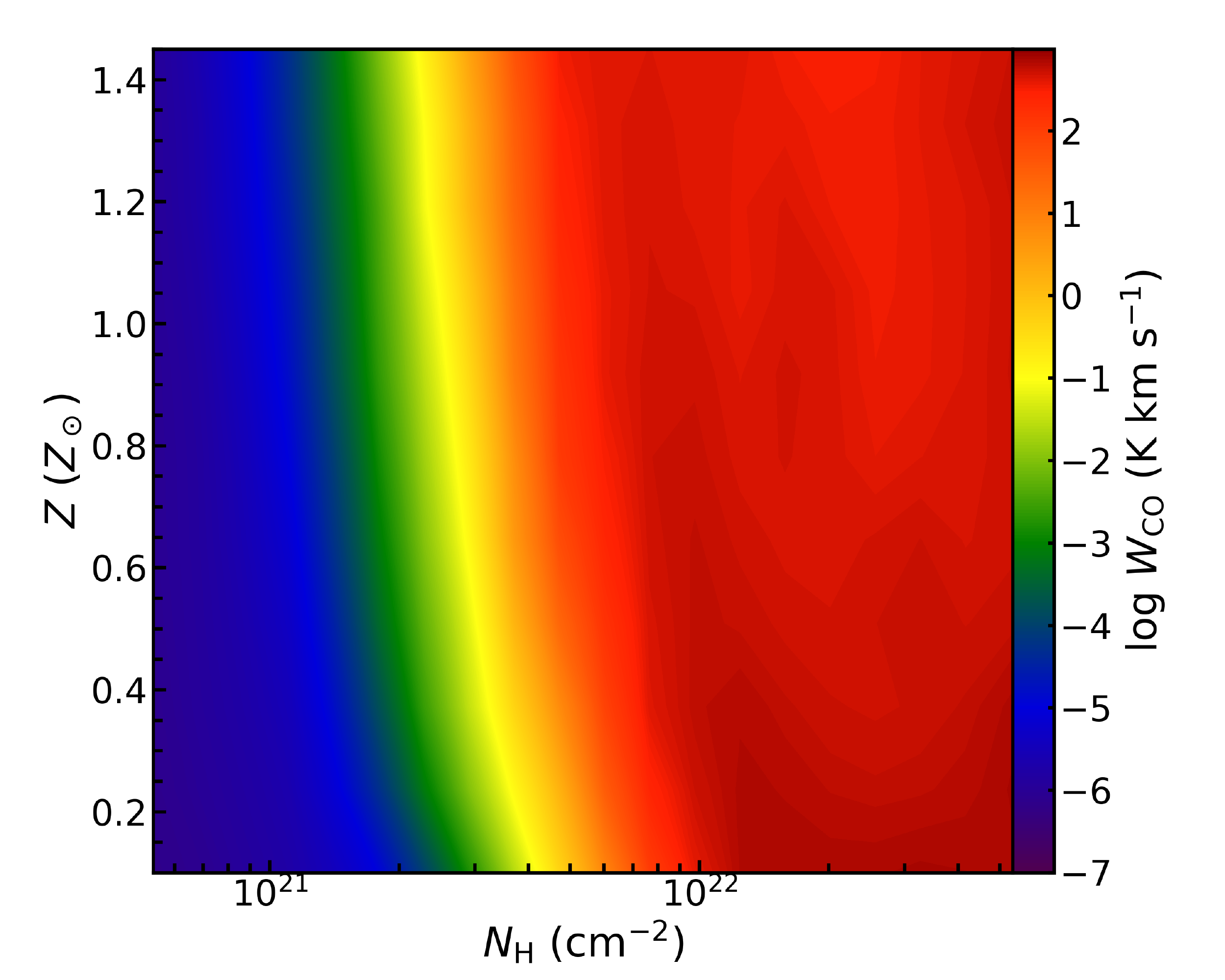}
\caption{The CO(1-0) intensity of molecular gas of Halo 401 against column density ($\NH$) and
  metallicity ($Z$). The intensity increases as $\NH$ and $Z$ increase due
  to stronger dust shielding.
  \label{fig:dark1}}
\end{figure}

In Figure~\ref{fig:dark1}, we show the CO(1-0) intensity of
molecular gas of an example galaxy (the $z=0$ snapshot of model Halo
401) as a function of cloud column density ($\NH$) and gas
metallicity (note, Halo 352 has a similar result).  
We show these two variables as they are critical, as a
combination, to shield CO from photodissociating Lyman-Werner band
photons. It is clear from that the CO-dark factor
depends strongly on $\NH$, which varies by several orders of magnitude
in our simulations, and that it mildly depends on $Z$.  One
requires column densities of $\sim 3 \times 10^{21}$~cm$^{-2}$ before
molecular clouds are CO-bright.  Another way of
saying this is that strongly CO-dark gas is typically relatively low
column density.

\begin{figure*}[]
\includegraphics[width = 1.0 \textwidth]{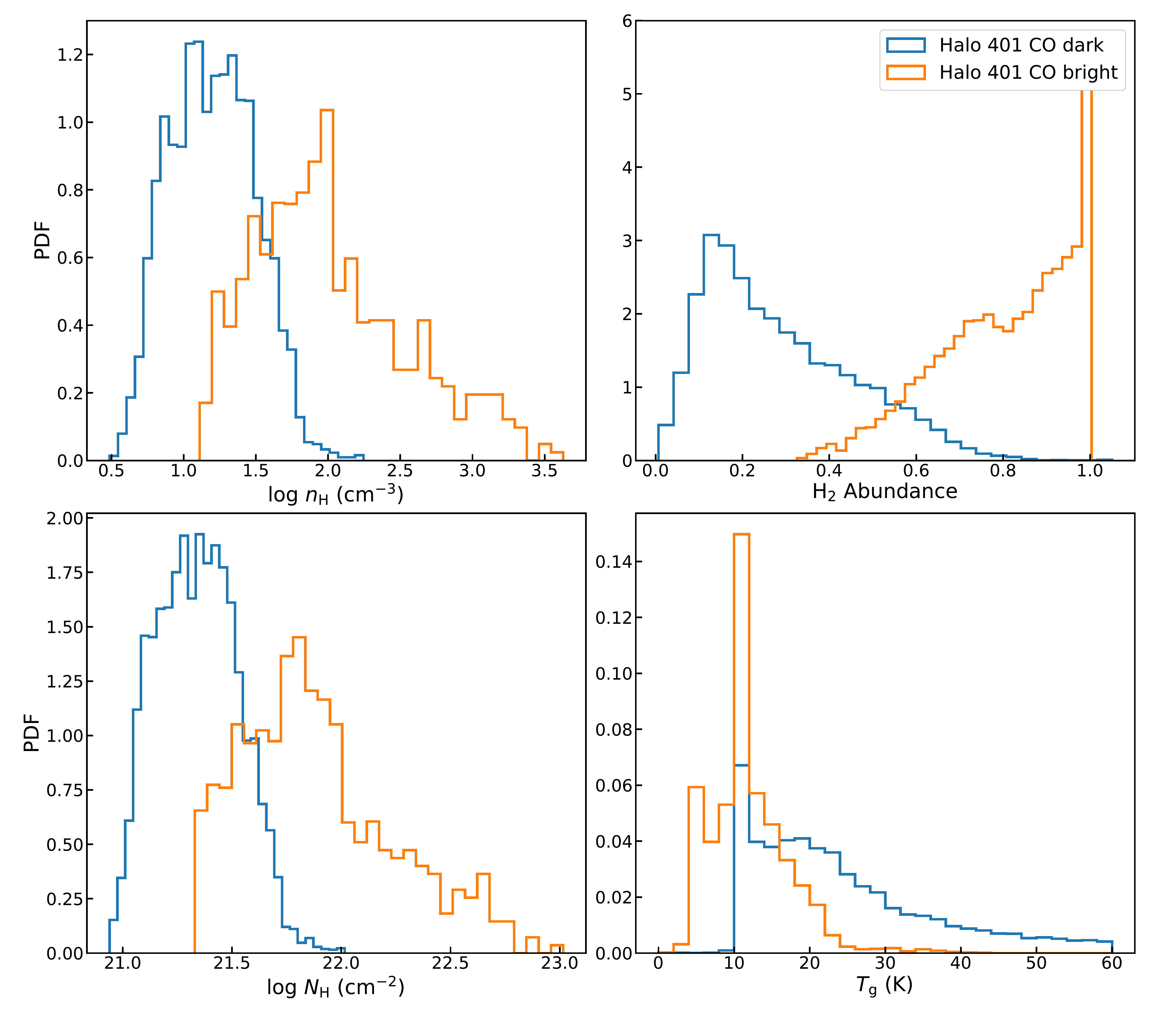}
\caption{PDFs of gas density ($\nH$; top left panel), $\Hm$ abundances
  (top right panel), column density ($N_{\rm H}$; bottom left panel)
  and gas kinetic temperature ($T_{\rm g}$, bottom right panel) for
  CO-dark (blue) and CO-bright (orange) gas.  CO-bright gas
  is principally highly extincted ($\NH \ga 60$ $M_\odot$~pc$^{-2}$),
  cold ($T_{\rm g} \la$10~K), dense gas ($\nH \ga  50\ cm^{-3}$) with
  extremely high ($\ga 60\%$) $\Hm$ fractions across the whole cloud,
  while the CO-dark counterpart is mainly diffuse gas.
		\label{fig:dark2}}
\end{figure*}

In Figure~\ref{fig:dark2}, we show the probability distribution
functions of gas density ($n_H$), $\Hm$ abundances, column density ($\NH$) and 
kinetic temperature ($T_{\rm g}$) for CO-dark and
CO-bright molecular-dominated clouds. CO-dark gas is typically relatively diffuse 
(i.e. low volume or column density), warm ($>10$ K) and
comprised of gas with relatively large H~\textsc{i} mass
fractions.

Taken together, these results paint a picture in which CO-dark
molecular gas is typically lower column density, lower density, and in
clouds with larger H~\textsc{i} fractions overall than CO-bright gas.  CO-bright
gas, on the other hand, is typically quite dense ($\nH \ga 50$
cm$^{-3}$), cold ($T_{\rm g} \sim$10~K), and highly extinct 
($\NH \ga 3 \times 10^{21}$~cm$^{-2}$).

\subsection{What Is the Best Method for Tracing CO-dark Gas?}
\label{sec:t}
Having established both the quantity and physical properties of
CO-dark gas, we now ask what the best method for tracing CO-dark gas
is.

\begin{figure*}[]
\includegraphics[width = 1.0 \textwidth]{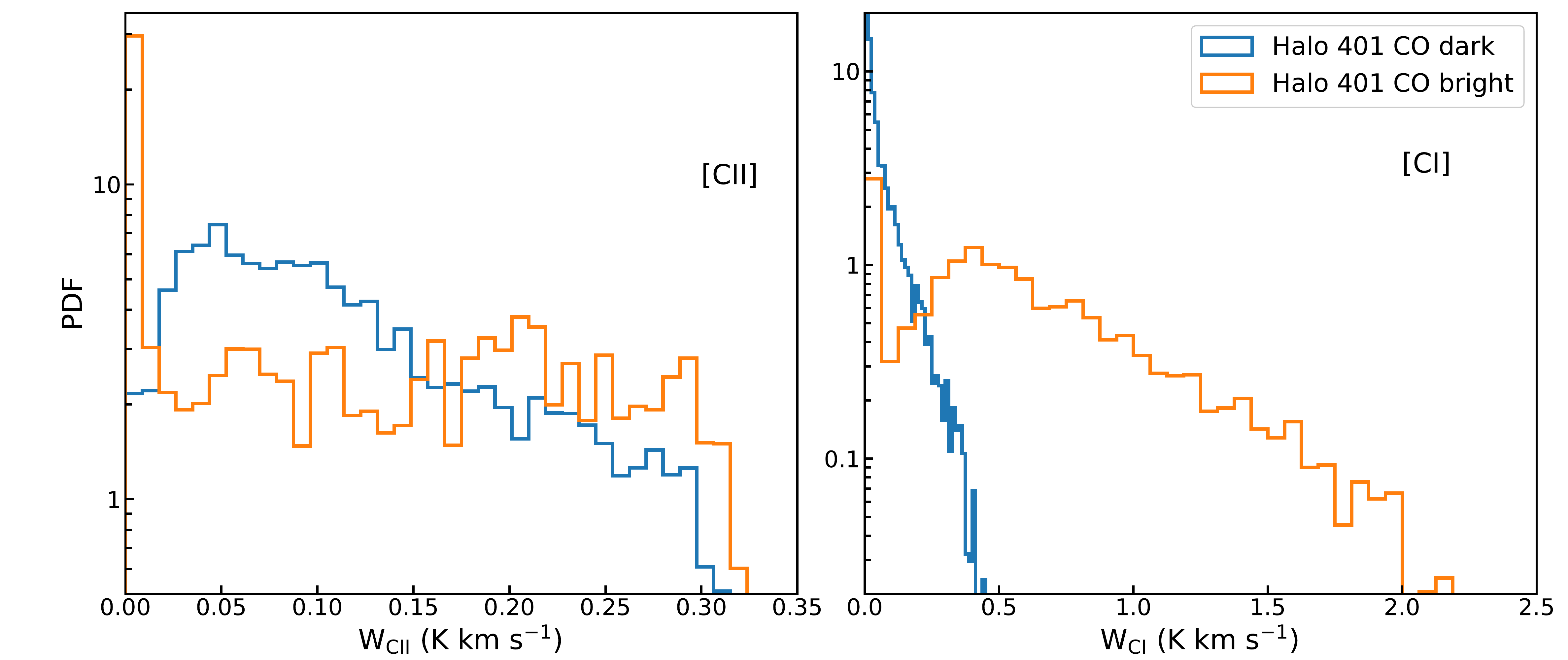}
\caption{PDFs of Intensities of [C~\textsc{ii}]
  (left panel) and [C~\textsc{i}] (right panel) emission from CO-dark (blue) and
  CO-bright (orange) gas.  Strong emission of both [C~\textsc{ii}] and [C~\textsc{i}]
  principally comes from CO-dark gas, while the emission is weak in
  CO-bright gas, especially for [C~\textsc{ii}]. Note, the line intensity is
  approximately proportional to $\nu^3 T_b$ in typical molecular
  clouds, where $\nu$ is line center frequency and $T_b$ is brightness
  temperature. Since $\nu^3_{\rm [C~\textsc{ii}]}/\nu^3_{\rm [C~\textsc{i}]}$ = 57.6,
  [C~\textsc{ii}] is overall more luminous than [C~\textsc{i}] in CO dark gas.
     \label{fig:dark3}}
\end{figure*}

First, we want to know how luminous alternative tracers are in
molecular gas.  We show the [C~\textsc{ii}] and [C~\textsc{i}] intensity of CO-dark gas
and compare them to the CO-bright counterpart in
Figure~\ref{fig:dark3}.  The [C~\textsc{ii}] intensity is brightest in the
CO-dark gas, and negligible in CO-bright gas.  At the same time, the
[C~\textsc{i}] emission, while naturally brighter in the CO-dark gas, is also
relatively bright even in the CO-bright gas. 
Note, the line intensity is approximately proportional to 
$\nu^3 T_b$ in typical molecular clouds, where $\nu$ is line center frequency
and $T_b$ is brightness temperature. Since $\nu^3_{\rm [C~\textsc{ii}]}/\nu^3_{\rm [C~\textsc{i}]}=57.6$,
[C~\textsc{ii}] has overall larger intensity than [C~\textsc{i}] in CO dark gas
though $T_{b,\mathrm{[C~\textsc{i}]}}$ is about twice as large as $T_{b,\mathrm{[C~\textsc{ii}]}}$.

We notice that, though [C~\textsc{ii}] emission
dominates in most of the CO-dark gas, [C~\textsc{i}] should also be strong enough to trace
CO-dark gas (more than $\nu^3_{\rm [C~\textsc{i}]}/\nu^3_{\rm CO(1-0)}=78$ 
times as luminous as CO(1-0) with $\WCO = 0.1$~K-km/s in general).
Besides, [C~\textsc{ii}] emission is generally too weak ($<0.05$~K~km~s$^{^-1}$) 
to trace CO-bright gas while strong in the
CO-dark counterpart. On the other hand, [C~\textsc{i}] emission is generally 
strong enough in CO-dark gas, while a range of intensities are found in 
CO-bright gas. This shows that [C~\textsc{i}] is capable of effectively tracing
molecular gas with a wider range of properties than [C~\textsc{ii}].

After establishing the potentially effective tracers for CO dark gas, we now ask how we
can use these tracers to determine the mass of underlying molecular
gas.  In other words, what are the [C~\textsc{i}], [C~\textsc{ii}] and CO(1-0) conversion
factors for CO bright and CO dark molecular gas?

\begin{figure}[]
\centering
\includegraphics[width = 0.5 \textwidth]{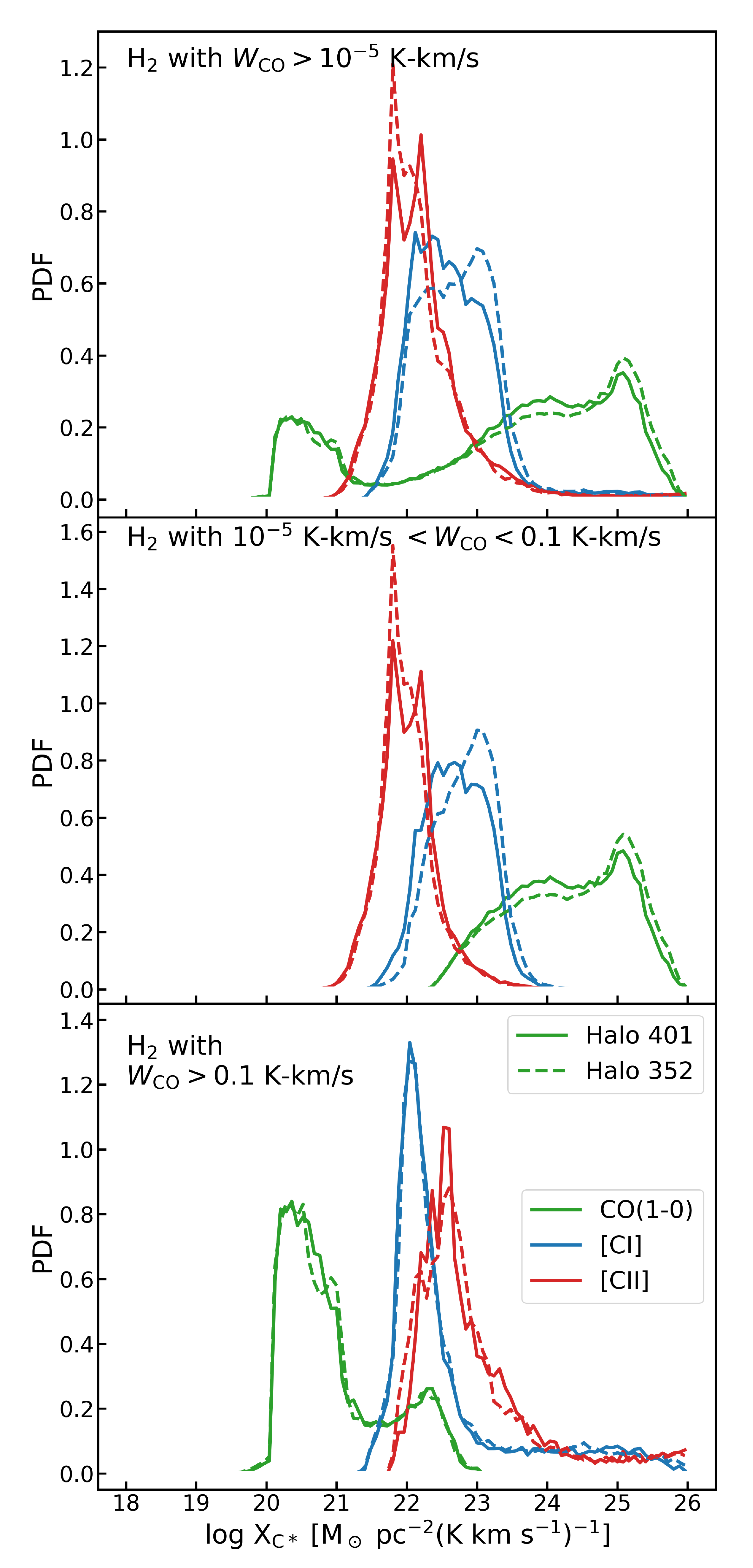}
\caption{ $\XCO$,
  $\XCI$ and $\XCII$ PDFs for Halo 401 and Halo 352.  Top panel: PDFs
  for all $\Hm$ with $\WCO>10^{-5}$~K-km/s; we can see two bumps for CO(1-0) 
  and one bump for [C~\textsc{i}] and [C~\textsc{ii}]. The larger bump of the 
  $\XCO$ PDF mainly consists of CO-dark gas while the smaller CO-bright counterpart.  
  Middle panel: PDFs for CO-dark gas (see \S~\ref{sec:def}) for the definition. 
  Bottom panel: PDFs for CO-bright gas. As is shown, the PDFs of $\XCI$ and
  $\XCII$ are much tighter than $\XCO$, implying weaker dependence on
  secondary parameters. [C~\textsc{i}] and [C~\textsc{ii}] works better for CO-dark gas 
  while the bimodal feature of $\XCO$ PDF complicates the interpretation of CO(1-0). 
 $\XCI$, rather than $\XCII$, remains stable over a wide range of physical properties.
\label{fig:t4-1}}
\end{figure}

In Figure~\ref{fig:t4-1}, we evaluate the
PDFs of the conversion factors for all $\Hm$ with $\WCO>10^{-5}$ K-km/s 
(see \S~\ref{sec:def};top), CO-dark $\Hm$ (middle), and CO-bright $\Hm$ (bottom);
a significant amount of molecular gas is traced by all three tracers over all gas 
within central galaxies of Halo 401 and 352.
 
From the top panel we notice that there are two bumps for the $\XCO$ PDF, revealing 
the impact of metal column density distribution and other secondary parameters. 
At the same time, distributions of $\XCI$ and $\XCII$ (especially $\XCI$) are relatively 
narrow with only one peak. The narrow distribution indicates that these two tracers 
are less sensitive to secondary parameters such as excitation, temperature, ISRF, and 
metallicity. Therefore these can reliably be used to trace the molecular gas in $z \sim 0$ disc galaxies.

Limiting the sample to CO-dark gas by our definition in \S~\ref{sec:def}, we find that there is 
little change of $\XCI$ and $\XCII$. The bimodal feature of $\XCO$, hower, is 
shifted to a single bump. We therefore conclude that [C~\textsc{i}] and [C~\textsc{ii}] perform better 
in tracing CO-dark molecular gas.

Shifting from CO-dark gas to the CO-bright counterpart, we find that the peak of $\XCI$ PDF roughly 
remains unchanged and that the PDF becomes tighter. The peak of $\XCII$ PDF, however, increases 
about 1 dex and deviates from the peak of $\XCII$ PDF of all $\Hm$. This shows [C~\textsc{i}], 
rather than [C~\textsc{ii}], remains a stable tracer over a wide range of physical properties.

\section{Discussion}
\label{sec:d}
\subsection{How Much H$_2$ Is Traced at Different Intensity Thresholds?}
\label{section:threshold_intensity}

\begin{figure}[]
\centering
\includegraphics[width = 0.5 \textwidth]{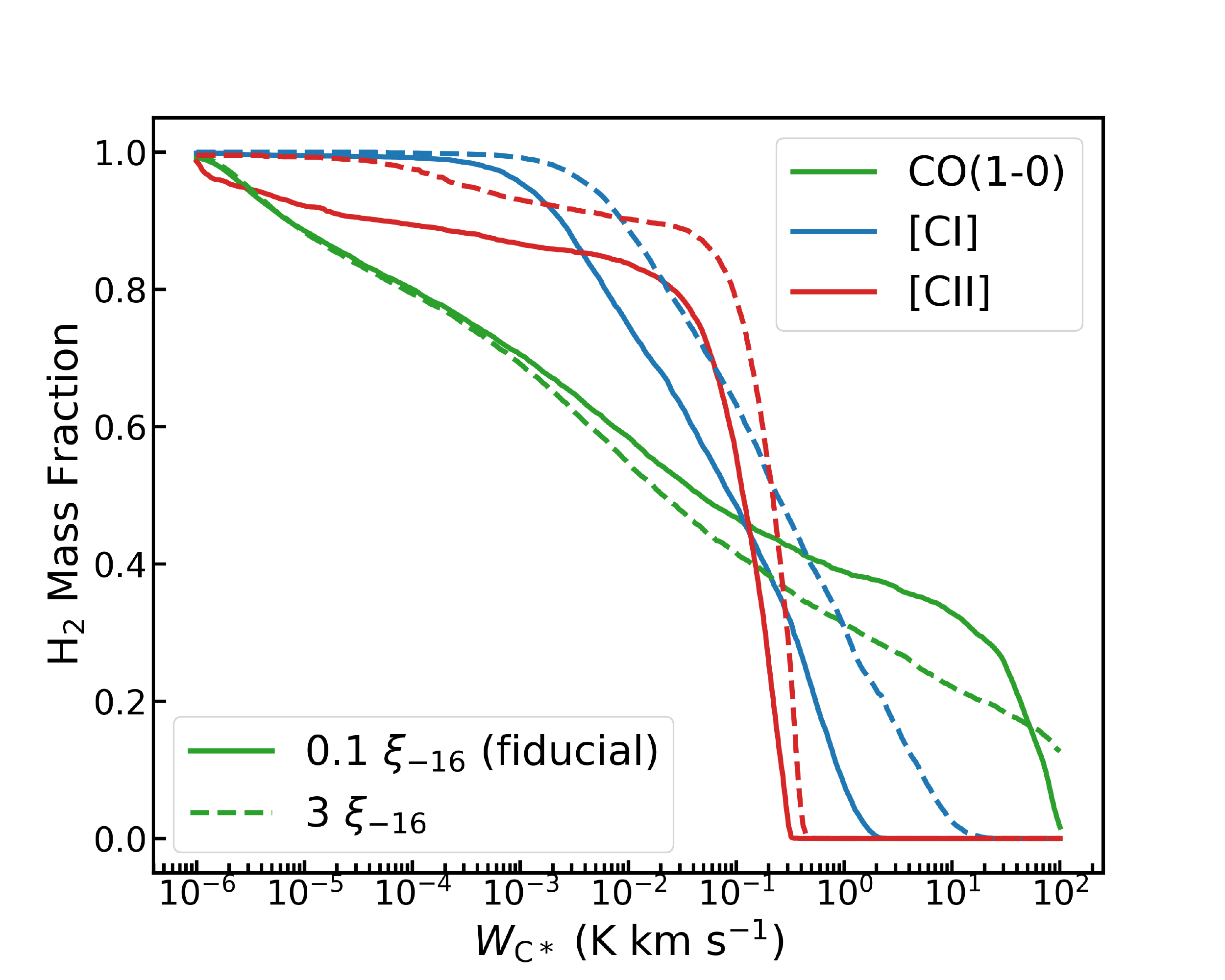}
\caption{ $\Hm$ mass fraction traced by CO (1-0), [C~\textsc{i}] and [C~\textsc{ii}] as
  functions of intensity threshold (above which the emission can be
  detected) for Halo 401. At higher intensity threshold, the
  traced molecular gas is dominated by CO (1-0). 
  however, CO (1-0) misses much of the molecular gas even with a very low threshold, and
  [C~\textsc{i}] and [C~\textsc{ii}] take over. For example, at $W_{\rm CO} \la 0.1$ K-km/s,
  CO(1-0) only traces a much smaller amount of $\Hm$ mass, compared to [C~\textsc{i}] and [C~\textsc{ii}].  As shown by the
  dashed lines, where we artificially increase our cosmic ray
  ionization rate by a factor $30$, this result is sensitive to the assumed cosmic ray ionization rate (futher discussed in \S~\ref{section:cosmic_rays}). 
  \label{fig:t1}}
\end{figure}
 We begin the discussion with Figure~\ref{fig:t1}, where we show how
 much molecular gas can be traced by different carbon tracers above a
 different observed intensity thresholds.  We use Halo 401 as an
 example, but note that Halo 352 has similar results.  We focus on the
 solid lines here, and defer discussion about the dashed lines (in
 which we vary the cosmic ray ionization rate) to
 \S~\ref{section:cosmic_rays}.

From the solid lines in Figure~\ref{fig:t1}, we see that at large
integrated intensity values (e.g. $W_{\rm C*} \sim 1$ K-km/s), the traced 
molecular gas is dominated by CO (1-0). In this regime, strong CO(1-0) comes from
copious CO residing in highly shielded and extremely $\Hm$-rich gas, where C~\textsc{i} and 
C$^{+}$ abundances are too low to generate strong emission. Even though strong [C~\textsc{ii}] and
 [C~\textsc{i}] emission comes from diffuse molecular gas, the intensity is not at the same level 
 as CO(1-0) from dense gas, due to the low gas density. At lower intensity thresholds,
 though, a much larger fraction of the gas is traced by [C~\textsc{i}] and [C~\textsc{ii}] while CO(1-0)
 misses a significant amount of $\Hm$ mass (e.g. $\sim 50\%$ missed at 
 $W_{\rm C*} \sim 0.1$ K-km/s), because most of the weak emission comes from diffuse molecular
 gas is CO-dark while [C~\textsc{i}] and [C~\textsc{ii}]-luminous (see also \S~\ref{section:results}).

\subsection{Sensitivity to Cosmic Ray Ionization}
\label{section:cosmic_rays}

The cosmic ray ionization rate in galaxies is uncertain. Observations
on diffuse molecular clouds
\citep[e.g.][]{wolfire10a,indriolo15a,neufeld10a,neufeld17a}
and dense ones \citep{mccall99a} within the Milky Way suggest rates
that range from $\xi\sim$0.1--3$\times$10$^{-16}$~s$^{-1}$.  Because
an increased cosmic ray ionization rate can increase the fraction of
dark gas (mainly via ionization of He$\rightarrow$He$^+$, and a
subsequent $2$ body reaction with CO), we therefore test the
sensitivity of our results to the assumed value of $\xi$.  Here, we
adopt a relatively extreme normalization of $\xi=$3$\times$10$^{-16}$
s$^{-1}$ for SFR$=1$ $\msun$/yr and compare the results to the
fiducial case of $\xi\sim0.1\times$10$^{-16}$ s$^{-1}$.

We return to Figure~\ref{fig:t1}, but now highlight the dashed lines,
in which we have assumed the larger ionization rate
$\xi\sim3\times$10$^{-16}$ s$^{-1}$.  The mass fraction of molecular
gas traced by [C~\textsc{ii}] and [C~\textsc{i}] above specific intensity thresholds
increases as CR ionization rate increases, while CO(1-0) traces less
molecular gas. This is a result of enhanced [C~\textsc{ii}] and [C~\textsc{i}] abundances,
as more CO is effectively destroyed by CR. The overall trend as
discussed in \S~\ref{section:threshold_intensity} remains unchanged,
but the enhanced CR ionization potentially make [C~\textsc{ii}] and [C~\textsc{i}] a
better tracer of molecular gas.

\begin{figure}[]
\centering
\includegraphics[width = 0.5 \textwidth]{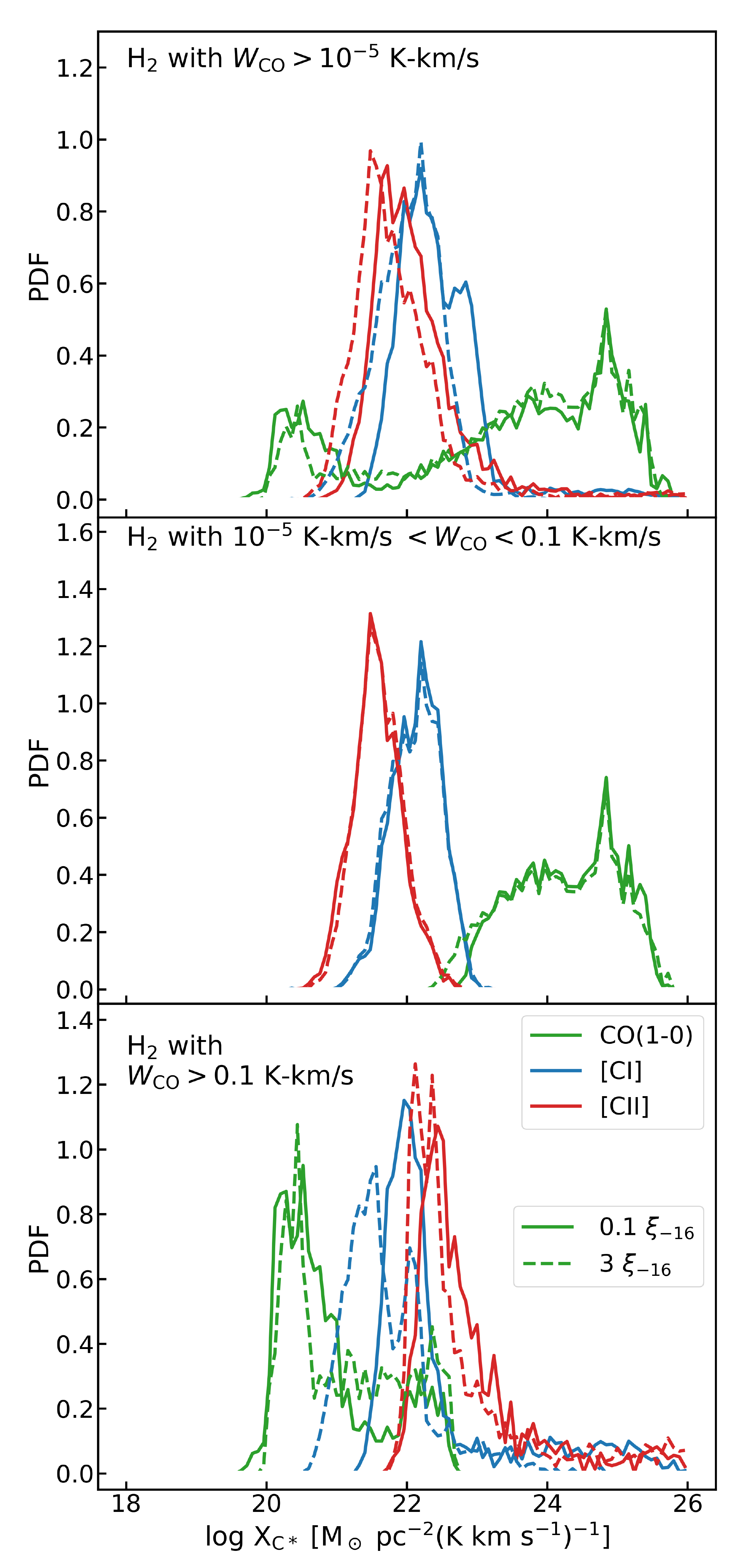}
\caption{ $\XCO$,
  $\XCI$ and $\XCII$ PDFs of Halo 401 for different normalization of cosmic ray ionization rates.  Top panel: PDFs
  for all molecular gas with $\WCO>10^{-5}$~K-km/s.  Middle panel: PDFs
  for CO-dark molecular gas.  Bottom panel: PDFs
  for CO-bright gas molecular gas.}
\label{fig:d1}
\end{figure}

We additionally check the effect of the normalization of CR ionization
rate on the conversion factors, shown in Figure \ref{fig:d1}.  
We can see that CO(1-0) performs even worse as a CO-bright $\Hm$
tracer as the PDF of $\XCO$ shows more power toward high $X$-factors
(again, a result of increased dissociation of CO). 
[C~\textsc{i}] and [C~\textsc{ii}] still perform well for CO-dark molecular gas.
[C~\textsc{ii}] still fails to trace part of diffuse gas as indicated by the 
shifting of the peak from top panel to bottom panel. Even if [C~\textsc{i}] 
is a still stable tracer of molecular gas with a wide range of physical properties, 
we notice, however, that the PDF of $\XCI$ becomes $\sim 1$ dex wider and 
the median is $\sim 0.5$ dex smaller compared to the fiducial case. 
This said, while $\XCI$ exhibits a relatively 
narrow distribution in values and remains stable across a wide range of physical 
properties,  we caution that the typical value and the dispersion are relatively 
sensitive to the assumed normalization of the CR ionization rate. 

We note, though, that these results may change as a result of the
choice of chemical network. For example, \citet{gong17a} find that
the NL99+GC network may overestimate the effectiveness of
cosmic rays in destroying CO, because it does not include
grain-assisted recombination.  This would suggest that the results
presented here are best thought of as an upper limit as to the
impact of cosmic rays. For comparison, in the Appendix we
implement the \citet{gong17a} chemical reaction network, and explore
the impact of of the cosmic ray ionization rate on our results.  As
we demonstrate in Figure~\ref{fig:a2}, the usage of the Gong et al. network 
does not change our overall conclusions, but the PDFs of $X$-factors 
are less sensitive to the assumed CR ionization rate, as expected.

\subsection{Comparing to Other Theoretical Work}
\label{section:compare}
Previous theoretical work \citep[e.g.][]{smith14a,glover16a,gong18a} on
quantifying CO-dark molecular gas and alternative molecule tracers has
typically been limited to studies of individual clouds or patches
extracted from isolated galaxies. These studies define
molecular gas as ``CO-dark" if $\WCO < 0.1$~K~km~s$^{-1}$. 
In particular, using a simulation of a patch of idealized Milky Way-like galaxy with
a simplified on-the-fly chemical network, \citet{smith14a} found 
$f_{\rm DG}\sim 42\%$. In their simulation, a significant amount of CO-dark gas is
located in long filaments which have low extinctions. They also show 
that the value of $f_{\rm DG}$ increase with increasing ISRF strength. Similarly, 
by post-processing 3D magnetohydrodynamics simulations of a kpc patch of galactic disks
with solar neighborhood conditions, Meanwhile, \citet{gong18a} find that $f_{\rm DG}$ 
is $26$ -- $79\%$, with the value of $f_{\rm DG}$ correlating with 
 the average extinction of the simulated patch. \citet{gong18a} also downgrade
the spatial resolution of their synthetic CO maps and column density maps to various
observational beam sizes, and find that in typical Galactic clouds,
$f_{\rm DG}$ decreases with the increasing beam size because CO
emission is smoothed out. \citet{gong18a} estimate $f_{\rm DG}\sim 50\%$ 
at the beam size $\sim 30$ pc which is similar to the minimum
smoothing length of gas particles in the central galaxies of our zoom
simulations at $z\sim 0$. Overall, our model gives a result
consistent with previous work quantifying CO-dark molecular gas
 when using apples-to-apples metrics.

\citet{offner14a} and \citet{glover15a} compare [C~\textsc{i}] against CO emission
in simulations of a cloud-scale turbulent ISM.  Contrary to the canonical view
of C~\textsc{i} as a ``surface'' tracer from simple 1-D PDR models
\citep[e.g.][]{tielens85,hollenbach99}, their numerical studies show
that C~\textsc{i} can be prevalent in molecular clouds with the help of
turbulent ``clumping'' and turbulent diffusion. They point out several
advantages of C~\textsc{i} over CO, e.g. the column density regime of [C~\textsc{i}] where
the corresponding conversion factor $\XCI$ remains approximately constant
is larger, and $\XCI$ is less sensitive to secondary parameters.  Our simulations 
support these conclusions, and extend them to a far larger dynamic range in ISM conditions.  
We note, however, that our simulations lack the resolution to resolve the effect of clumping 
on the extinction of ISRF and the increased surface area exposed to ISRF in small scales.
Still, we do consider the effect of sub-grid clumping on collisional thermal and chemical processes, and a simulation with $\epsilon \sim 25$~pc (close to $\epsilon \sim 30$~pc of our simulations), would have minimal sub-grid clumping effect on volume and column densities \citep{dave16a}.

\section{Conclusions}
\label{sec:c}
Combining cosmological hydrodynamic zoom-in galaxy formation
simulations of $z\sim 0$ disc galaxies with
thermal-radiative-chemical equilibrium interstellar ISM calculations,
we investigate the utility of CO(1-0), [C~\textsc{i}] and [C~\textsc{ii}] emission as
molecular tracers in the environment of non-isolated galaxies. We
summarize our main findings as following.
\begin{itemize}
\item Most CO(1-0), [C~\textsc{i}] and [C~\textsc{ii}] emission comes from molecular gas, 
but as we transition from molecular (CO 1-0) to atomic ([C~\textsc{i}]) to ionized ([C~\textsc{ii}]) carbon
emission, the fraction of emission that originates in molecular gas decreases, and the fraction that originates in atomic gas increases. 
\item We define CO-dark gas as molecular gas . This criterion
  effectively distinguishes dense molecular gas from diffuse molecular
  gas. Using our definition of CO-dark gas, we find that the CO-bright 
  portion of the ISM consists principally of highly shielded ($\NH
  \gtrsim200\ \msun\ \pc^{-2}$) gas with extremely high densities
  ($n_{\rm H}>300\cm^{-3}$) and $\Hm$ abundances ($>80\%$).  
 As a result, observations that focus principally on CO can miss significant 
 amounts of molecular gas.
 \item With our definition of CO-dark gas as the observationally motivated
 $W_{\rm CO} < 0.1$~K-km/s, the simulated disc galaxies have a
  significant amount ($\sim 53\%$) of CO-dark gas which can be traced by [C~\textsc{ii}] 
  and [C~\textsc{i}], emphasizing the importance of these tracers.
\item We show the PDFs of CO(1-0), [C~\textsc{i}] and [C~\textsc{ii}] conversion
  factors $\XCO$, $\XCI$ and $\XCII$ between observed line intensity and true H$_2$ 
  column density. Of these, $\XCI$ and $\XCII$ tend to have a tighter 
  distribution than $\XCO$, due to the sensitivity of CO (1-0) emission to secondary 
  parameters such as metallicity.
\item We find that [C~\textsc{i}] is the overall preferable tracer of molecular gas 
within a wide range of physical properties. $\XCI$ remains stable when shifted 
from CO-dark gas to CO-bright gas. [C~\textsc{i}] is more luminous than CO(1-0) 
in CO-dark gas and bright in CO-bright gas, allowing it to trace a larger amount of 
$\Hm$ mass.
\end{itemize}

\acknowledgements{Q.L. was funded by NSF grant AST-1724864.  D.N. acknowledges funding
from NSF grant AST-1724864, AST-1715206 and HST AR-13906.001 from the
Space Telescope Science Institute. M.R.K. acknowledges support from the 
Australian Research Council's Discovery Projects funding scheme, grant DP160100695.}

\begin{appendix}

In order to determine the sensitivity of our results to our choice of chemical network, in this appendix we repeat some of the key calculations in the paper using the \citet{gong17a} chemical network instead of our fiducial NL99+GC network.\\

\section{Chemical Structures of a Typical Cloud}
\begin{figure*}[]
\includegraphics[width = 1 \textwidth]{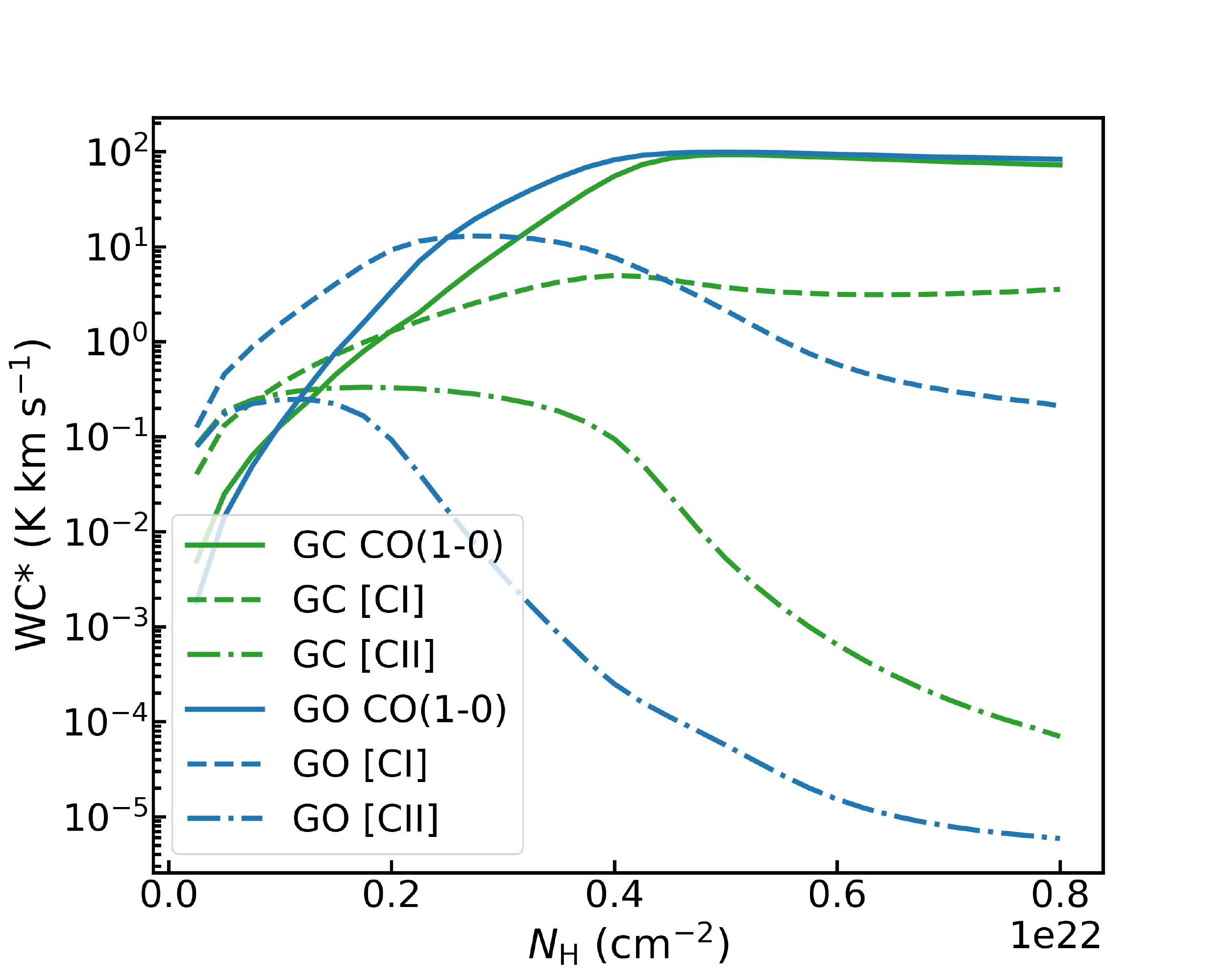}
\caption{CO(1-0) (solid line), [C~\textsc{i}] (dashed line) and [C~\textsc{ii}] (dotted line) intensity as a function of $\NH$ for a 
radially stratified gas particle with $\nH=105.87 \cm ^{-3}$, $\NH=149.26 \msun \ \pc ^{-2}$, $Z'=1.04$, 
and SFR$=1$ $\msun$/yr, using the \citet{gong17a} network (blue) and our fiducial NL99+GC network (green), respectively}
\label{fig:a1}
\end{figure*}

We show in Figure~\ref{fig:a1} the chemical structures (i.e. radial profiles) of a typical molecular cloud 
with $\nH=105.87 \cm ^{-3}$, $\NH=149.26 \msun \ \pc ^{-2}$, $Z'=1.04$, and SFR$=1$ $\msun$/yr.
We adopt here the CR ionization rate $\xi = 10^{-17}$~s$^{-1}$. We notice that the grain-catalyzed 
recombination incorporated in \citet{gong17a} network leads to a significantly higher [C~\textsc{i}] and CO(1-0) intensities in the outer layers ($\NH \leqslant40 \msun \ \pc ^{-2}$) where the ionization via CRs and energetic ISRF is strong.

\section{Conversion Factors of Three Carbon-based Tracers}
\begin{figure*}[]
\includegraphics[width = 1 \textwidth]{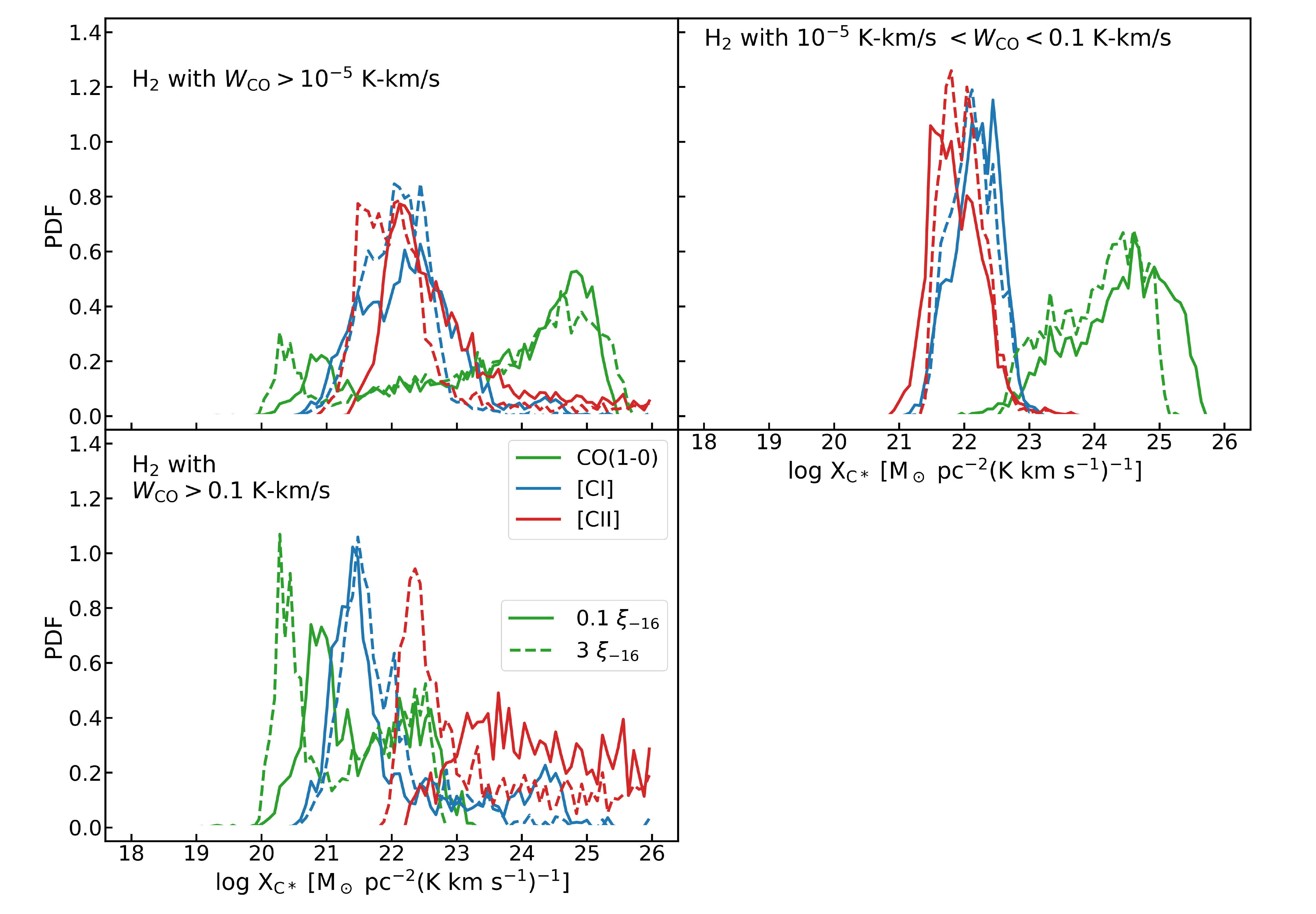}
\caption{ $\XCO$, $\XCI$ and $\XCII$ PDFs of Halo 401 for different 
cosmic ray ionization rates, computed using the
  \citet{gong17a} network. Top-left panel: PDFs
  for all molecular gas with $\WCO>10^{-5}$~K-km/s. Top-right panel: PDFs
  for CO-dark gas by our definition (see \S~\ref{sec:def}). 
  Bottom-left panel: PDFs for CO-bright gas.
\label{fig:a2}}
\end{figure*}

We show in Figure~\ref{fig:a2} the PDFs of $\XCO$, $\XCI$ and $\XCII$ of Halo 401 for two different cosmic ray ionization rates, using \citet{gong17a} networks. Comparing the result to the previous discussion in \S~\ref{section:cosmic_rays}, the inclusion of grain-catalyzed recombination does not change our overall conclusions. 

\section{Resolution Tests}
\begin{figure*}[]
\includegraphics[width = 1 \textwidth]{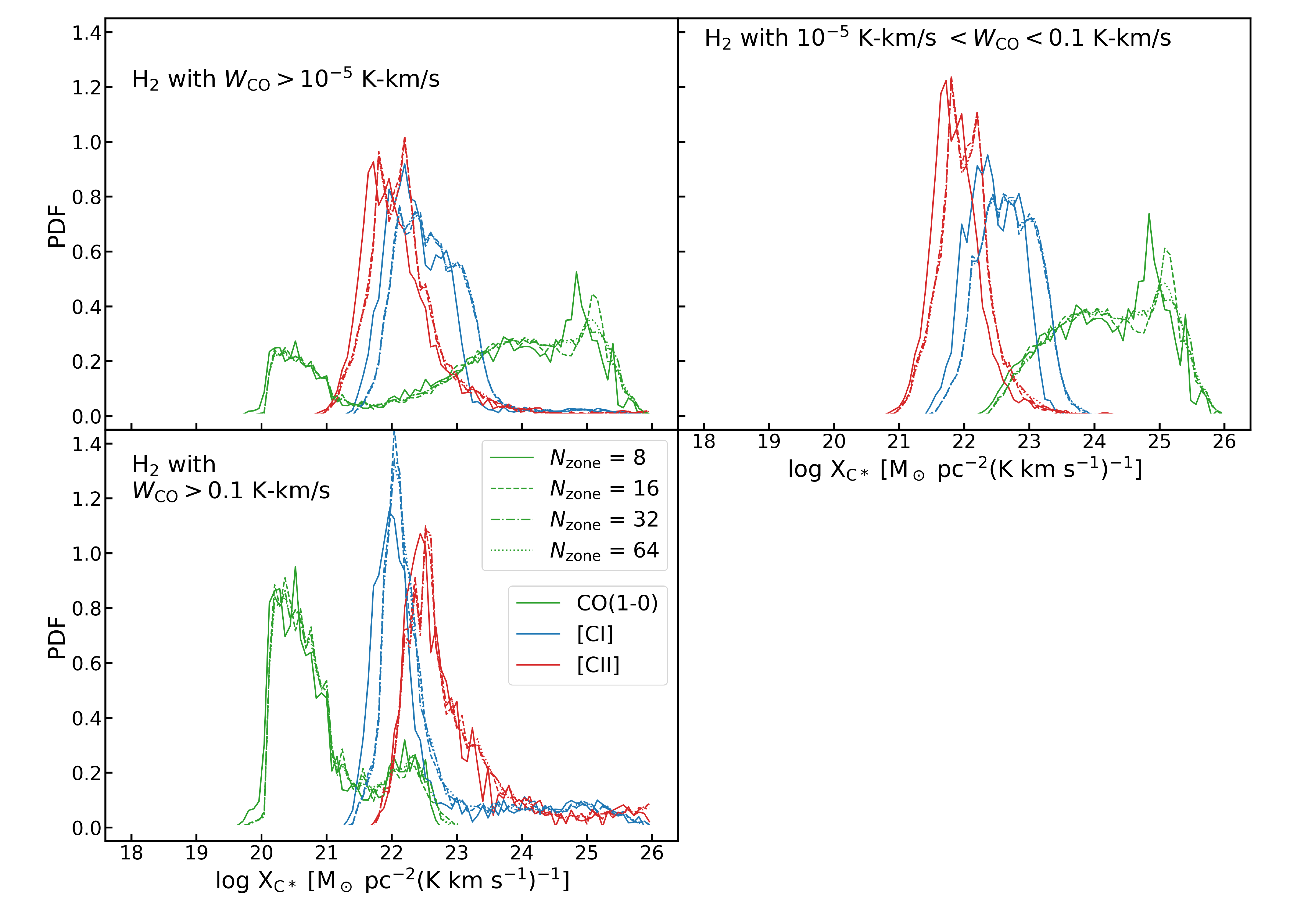}
\caption{Convergence tests show $\XCO$, $\XCI$ and $\XCII$ PDFs of Halo 401 for models with different 
number of layers ($N_{\rm zone}=8$, $16$, $32$ and $64$), computed using the
  NL99+GC network and fiducial cosmic ray ionization rates $\xi = 10^{-17}$
SFR/$M_\odot\ \mathrm{yr}^{-1}$ s$^{-1}$. Top-left panel: PDFs
  for all molecular gas with $\WCO>10^{-5}$~K-km/s. Top-right panel: PDFs
  for CO-dark gas by our definition (see \S~\ref{sec:def}). 
  Bottom-left panel: PDFs for CO-bright gas.
\label{fig:a3}}
\end{figure*}

We show in Figure~\ref{fig:a3} the PDFs of $\XCO$, $\XCI$ and $\XCII$ of Halo 401 for models 
with different number of layers ($N_{\rm zone}=8$, $16$, $32$ and $64$). The tests utilize a cruder grid, the spacing of which is 0.28,
0.4 dex, 0.4 dex, 0.3 dex for $Z'$, $\nH$, $\NH$ and SFR respectively. The differences are minor.

\end{appendix}

\bibliography{paper_refs}

\end{document}